%% file: Main.tex
\documentclass[iicol,pdflatex,sn-mathphys-num]{sn-jnl}

\usepackage{pgfplots}
\usepackage{graphicx}%
\usepackage{multirow}%
\usepackage{amsmath,amssymb,amsfonts}%
\usepackage{amsthm}%
\usepackage{bm}
\usepackage{mathrsfs}%
\usepackage[title]{appendix}%
\usepackage{xcolor}%
\usepackage{textcomp}%
\usepackage{manyfoot}%
\usepackage{booktabs}%
\usepackage{algorithm}%
\usepackage{algorithmicx}%
\usepackage{algpseudocode}%
\usepackage{listings}%
\usepackage{siunitx}
\usepackage{cleveref}
\usepackage{comment}
\usepackage{xcolor}
\usepackage{bm}
\usepackage{cuted}
\usepackage{stfloats}  

\usepackage{xcolor}
\usepackage{xspace}
\usepackage{subcaption}

\usepackage{lineno}[switch,displaymath]
\modulolinenumbers[5]

\theoremstyle{thmstyleone}%
\theoremstyle{thmstyletwo}%

\theoremstyle{thmstylethree}%
\input{commands}

\begin{document}

\title[Article Title]{The cross-sectional warping problem for hyperelastic beams: An efficient formulation in Voigt notation}


\author*[1]{\fnm{Juan C.} \sur{Alzate Cobo}} 
\email{alzate@cps.tu-darmstadt.de}

\author[1]{\fnm{Tobias} \sur{Henkels}} 

\author[1]{\fnm{Oliver} \sur{Weeger}}

\affil[1]{\orgdiv{Cyber-Physical Simulation}, \orgname{Technical University of Darmstadt}, \orgaddress{\street{Dolivostr. 15}, \city{Darmstadt}, \postcode{64293}, \country{Germany}}}


\abstract{Beam theory has traditionally been restricted to small elastic strains and rigid cross-sections. Relaxing these assumptions within closed-form analytical frameworks remains challenging. In contrast, the cross-sectional warping problem provides a computational approach that enables the derivation of general, nonlinear constitutive relations for beam models, thereby overcoming both limitations. In this work, we reinterpret the cross-sectional warping problem for hyperelastic beams and propose a fully material formulation in terms of the Green-Lagrange strain and the second Piola-Kirchhoff stress tensors. Owing to the symmetry of these tensors, the formulation can be expressed efficiently in Voigt notation and is thus particularly well-suited for straightforward numerical implementation. We demonstrate the validity of this alternative formulation in numerical examples, including the computation of the effective beam stiffness, for which we derive the sensitivities of the warping displacement. To promote reproducibility, we accompany this article with an open-access repository containing the isogeometric finite element implementation and all numerical examples presented herein, enabling other researchers to readily reproduce and build upon our results.}




\maketitle


\input{S1Introduction}

\input{S2BeamTheory}
\input{S3TheCSWP}

\input{S4IGA}

\input{S5Results}
\input{S6Conclusion}
\backmatter

\bmhead{Supplementary information}

An accompanying \href{https://github.com/CPShub/CSWP_Voigt}{GitHub repository} contains all numerical results and implementations presented in this work, enabling full reproducibility and providing access to the capabilities described in the paper.

\bmhead{Acknowledgements}
The authors acknowledge the financial support provided by the Deutsche Forschungsgemeinschaft (DFG, German Research Foundation, project number 460684687), as well as the Graduate School of Computational Engineering at TU Darmstadt. 



\noindent

%
%
%


\bibliography{sn-bibliography}

\end{document}

%% file: commands.tex
\newcommand{\e}[1]{\mathbf{e}_{#1}}

\newcommand{\mf}[1]{\mathbf{#1}}
\newcommand{\hatn}{\hat{\mathbf{n}}}
\newcommand{\hatm}{\hat{\mathbf{m}}}
\newcommand{\tb}{^{\text{b}}}
\newcommand{\R}{\mathbf{R}}

\newcommand{\T}{^{\mathrm{T}}}

\newcommand{\XiI}{\mu_3 \frac{2 x_1 x_2}{(x_1^2+x_2^2)^2}}
\newcommand{\XiII}{\mu_3 \frac{ x_2^2- x_1^2}{(x_1^2+x_2^2)^2}}

\newcommand{\lmlmp}{\bm{\lambda}} 
\newcommand{\mulmp}{\bm{\mu}}

\newcommand{\mx}{\{\mf x\}_\times}
\newcommand{\mxh}{\{\mf x_h\}_\times}

\newcommand{\kR}{\bm{\kappa}}
\newcommand{\epsR}{\bm{\varepsilon}}

\definecolor{CPSgreen}{RGB}{22,164,138}
\definecolor{CPSlightblue}{RGB}{104,143,198}
\definecolor{CPSdarkblue}{RGB}{67,83,132}
\definecolor{CPSgrey}{RGB}{204, 204, 204}
\definecolor{CPSorange}{RGB}{246,163,21}
\definecolor{CPSred}{RGB}{194,76,76}
\definecolor{CPSredtwo}{RGB}{160,0,0}
\definecolor{CPSgreentwo}{RGB}{0,128,128}


%% file: S1Introduction.tex
\section{Introduction}\label{Sec: Intro}

Beam-based structures are widely used in applications involving nonlinear and inelastic material behavior, including concrete reinforcements, soft robots, mechanical and multiphysical metamaterials, woven and knitted textiles, and more \cite{SeismeicFramesPriestley2000,SofRobotsRus2015,SofContinuuRobotsBurgner2015,xia_electrochemically_2019,StrainRateMetaMatjanbaz2020,karthikeyan_3D_TEGs_2023,weegerInelastic2022,weegerNonlinearMultiscaleModelling2018,ding4DRods3D2017}. However, the computational design and simulation of such structures using three-dimensional (3D) continuum models and their finite element discretizations are computationally intensive, and more efficient alternatives are therefore desirable. One such alternative is provided by one-dimensional (1D) models based on beam theory. Nevertheless, classical beam theory exhibits major limitations: it is restricted to small strains, linear elastic material behavior, and deformations in which the cross-section remains planar \cite{antman_nonlinearElasticity_2005,auricchio_Cosserat_2008,eugster2014foundations,meier_FEMBeamtheores_2019,mittelstedt_StructuralMechanics_2021a}.

Enhancing beam theory with nonlinear constitutive models is far from trivial. Recently, significant research has been made to extend classical beam theory to include finite elastic strains and hyperelastic materials \cite{choi_ExtensDirectors_2021,ignesti_hyperelasticBeams_2026}, plasticity and visco-elasticity at small strains \cite{smriti_finite_2021,smriti_thermoelastoplastic_2019,weeger_PlasticBeam_2022,ferri2023efficient}, chemo-elasticity \cite{parida_chem_el_Beam26}, and even large swelling expansions caused by anisotropic diffusion \cite{alzate_FSB_2025}. However, the aforementioned works are either based on additional degrees of freedom, on small elastic strains, on overly restrictive kinematics, or are even incomplete, as can be seen in the context of plasticity, for which yield surfaces and hardening laws in terms of stress resultants, i.e., in terms of forces and moments, are still a matter of active investigation \cite{herrnbock_YieldSurface_2021,herrnbock_two-scale_2023,gartnerBeamHardening2025}. Most importantly, a unifying approach for obtaining beam constitutive models which includes hyperelastic materials, inelastic effects such as swelling and visco-elasto-plasticity at both large and small strains, and that is simultaneously conformal with 3D continuum kinematics, i.e., that allows cross-sectional deformations, is largely missing from the literature.

In order to bridge the gap between computationally efficient yet kinematically restricted beam models and their kinematically general but numerically expensive 3D continuum counterparts, Arora et al. \cite{arora_computational_2019} introduced the cross-sectional warping problem (CSWP). The CSWP is based on the Helical Cauchy-Born rule (HCB) proposed by Kumar et al. \cite{kumar_HCBR_2016}, which reduces beam deformations to a family of helical configurations by imposing a uniform strain field along the beam centerline. These configurations depend solely on the six strain measures of beam theory. Owing to the uniformity of the strain field along the beam axis, the problem reduces to the cross-section and can be formulated as a boundary value problem (BVP). Its solution yields cross-sectional deformations, from which resultant forces, moments, and stiffnesses can be obtained for arbitrary strain states. By modeling the cross-section as a two-dimensional (2D) continuum with three displacement degrees of freedom, the CSWP captures both in-plane and out-of-plane deformations and accommodates general hyperelastic material behavior. Typically, the resulting BVP is discretized and solved using an isoparametric finite element formulation \cite{arora_computational_2019}.

In recent years, the CSWP and variations thereof have been used for the determination of a variety of nonlinear beam constitutive relations and cross-sectional effects. For example, Singh et al. \cite{singh_CSWP_SurfaceEnergy_2022} enhanced it with surface energy, Herrnböck et al. \cite{herrnbock_YieldSurface_2021} employed it for obtaining yield surfaces in terms of beam measures and then enhanced the model to include hardening \cite{herrnbock_two-scale_2023}, Vinayak et al. \cite{vinayak_CSWP_Plastic_2023} used it for modeling anisotropic plasticity, Kumar et al. \cite{kumarStrips2024} used it as a foundation for obtaining nonlinear elastic constitutive models in strips, and, most recently, Kumar et al. \cite{kumar_growth-CSWP_2026} extended it to include cross-sectional growth. Additionally, the CSWP has been embedded in FE\textsuperscript{2}-type concurrent multiscale frameworks in elasto-plastic beams with hardening \cite{herrnbock_two-scale_2023,herrnbock_PhdThesis_2023} and for hyperelastic, shear-rigid beams \cite{le_clezio_2Scale_2023}, where it replaces an explicit constitutive model. Furthermore, it has been used for data generation in the context of learning hyperelastic beam constitutive models using thermodynamically consistent neural networks \cite{schommartz_PANN_CSWP_2025, le_clezio_NN_2024}. 

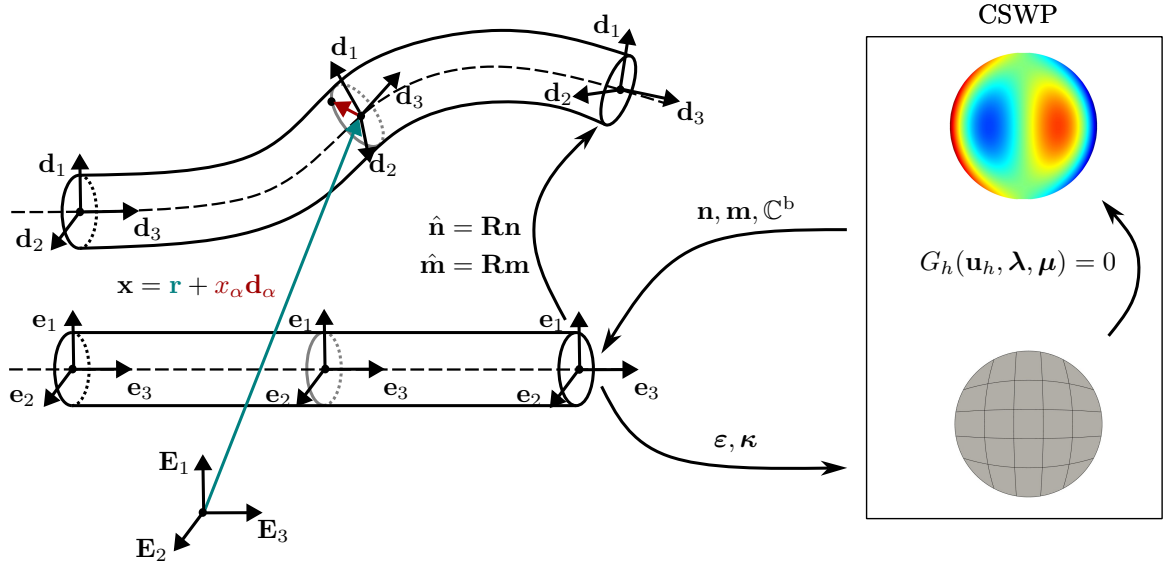
\begin{figure*}[!t]
    \centering
    \input{Sketch}
    \caption{Deformation map (left) of a geometrically nonlinear beam and schematic representation of the cross-sectional warping problem as a means of obtaining nonlinear constitutive relations suitable for beam theory (right)} 
    \label{fig: Beam Kinematics}
\end{figure*}

One drawback of the CSWP formulation proposed by Arora et al. \cite{arora_computational_2019}, and its subsequent extensions \cite{singh_CSWP_SurfaceEnergy_2022,vinayak_CSWP_Plastic_2023,herrnbock_YieldSurface_2021,herrnbock_two-scale_2023,kumar_growth-CSWP_2026}, lies in its energy-conjugate formulation in terms of the first Piola-Kirchhoff (PK1) stress tensor and the deformation gradient. In particular, its numerical implementation requires the manipulation of full, non-symmetric second-order stress and strain tensors and entails a complex linearization involving fourth-order constitutive tensors. In contrast, standard formulations in finite strain continuum mechanics are typically expressed in a fully material framework based on the symmetric second Piola-Kirchhoff (PK2) stress tensor and the right Cauchy-Green or Green-Lagrange (GL) strain tensor \cite{wriggers_NLFEM_2008}. As in linear elasticity, this fully material formulation enables the use of an efficient Voigt matrix-vector representation of symmetric second- and fourth-order tensors, thereby significantly simplifying implementation and reducing computational cost. 

Motivated by these advantages, the present work introduces a reformulation of the CSWP for shear-deformable, hyperelastic 3D beams based on a fully material framework in Voigt notation. This approach yields a more concise theoretical description and a computationally more efficient numerical implementation.
The highlights of this work are:
\begin{enumerate}
    \item Energy-conjugate formulation of the CSWP for hyperelastic, shear-deformable 3D beams in terms of PK2 stress and GL strain.
    \item Derivation and use of the strain--displacement differential operators ($\mf B$ and $\mathfrak{B}$) for solving the CSWP in Voigt notation and for determining the tangent stiffness of the beam.
    \item Detailed description of the numerical implementation. 
    \item Numerical verification of the proposed fully material ``PK2 formulation'' through comparison with a ``PK1 formulation'' based on \cite{arora_computational_2019} and with established results from the literature.
    \item Open-source code publication, based on the \texttt{MATLAB} isogeometric analysis library \texttt{NLIGA} \cite{du_nliga_2020}, at a GitHub repository\footnote{\url{https://github.com/CPShub/CSWP_Voigt}},
    including all results from both the PK1 and the PK2 formulation.
\end{enumerate}

The remainder of this article is structured as follows: 
In \cref{Sec. Beam theory}, a brief introduction to the geometrically exact 3D beam theory is given.
This is followed by the derivation of the cross-sectional warping problem in a novel, fully material formulation in \cref{Sec. CSWP}.
Then, in \cref{Sec. IGA}, an isogeometric finite element discretization of the CSWP is detailed, including the derivation of the $\mf B$ and $\mathfrak{B}$ strain--displacement operators as well as their use in the determination of the tangent stiffness matrix of the beam. 
\Cref{Sec. Results} provides a detailed verification of the approach, 
followed by a conclusion in \cref{Sec. Concl}. 

\medskip

\textbf{Notation:} Unless stated otherwise, indices with Roman letters $\bullet_i$ take values $i=1,2,3$, whereas indices with Greek letters $\bullet_\alpha$ take values $\alpha=1,2$. Furthermore, in accordance with standard index notation, summation over repeated indices is implied. Partial derivatives are denoted by a comma, such that $\partial_i \bullet = \bullet_{,i}$.

%% file: Sketch.tex
\begin{tikzpicture}[xscale=1,yscale=1]
\useasboundingbox(0,0) rectangle (170mm,80mm);
\begin{scope}[shift={(-5mm,0)}]
\node[draw=none,fill=none] at (0,0){\includegraphics[scale=1,bb=0 0 0 0]{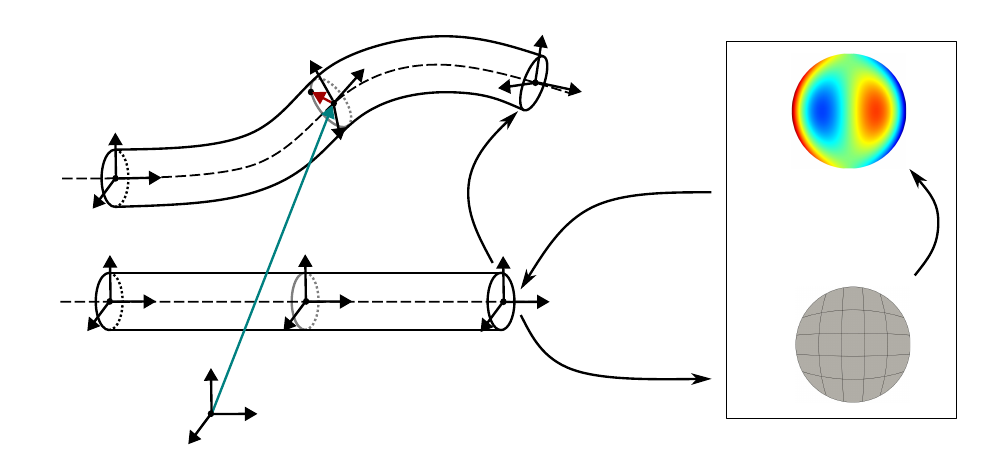}};
\end{scope}

 \begin{scope}
     \node at (24mm,5mm) [] {$\mf E_2$};
     \node at (40mm,7.5mm) [] {$\mf E_3$};
     \node at (27mm,16.5mm) [] {$\mf E_1$};
 \end{scope}
  \begin{scope}[shift={(-17.5mm,18.5mm)}]
     \node at (24.5mm,6.6mm) [] {$\mf e_2$};
     \node at (40mm,7.5mm) [] {$\mf e_3$};
     \node at (27.5mm,16.5mm) [] {$\mf e_1$};
 \end{scope}

  \begin{scope}[shift={(16.2mm,18.5mm)}]
     \node at (24.5mm,6.6mm) [] {$\mf e_2$};
     \node at (40mm,7.5mm) [] {$\mf e_3$};
     \node at (27.5mm,16.5mm) [] {$\mf e_1$};
 \end{scope}

  \begin{scope}[shift={(49.2mm,18.5mm)}]
     \node at (24.5mm,6.6mm) [] {$\mf e_2$};
     \node at (40mm,7.5mm) [] {$\mf e_3$};
     \node at (27.5mm,16.5mm) [] {$\mf e_1$};
 \end{scope}
 \node at (138mm, 76mm) [] {CSWP};

  \begin{scope}[shift={(-16.5mm,39.5mm)}]
     \node at (24.5mm,6.6mm) [] {$\mf d_2$};
     \node at (40mm,8mm) [] {$\mf d_3$};
     \node at (27.5mm,16.5mm) [] {$\mf d_1$};
 \end{scope}
  \begin{scope}[shift={(22mm,55mm)}]
     \node at (32.5mm,1mm) [] {$\mf d_2$};
     \node at (36mm,10mm) [] {$\mf d_3$};
     \node at (27.5mm,16.5mm) [] {$\mf d_1$};
 \end{scope}
  \begin{scope}[shift={(55mm,55mm)}]
     \node at (22.5mm,10.5mm) [] {$\mf d_2$};
     \node at (40mm,8mm) [] {$\mf d_3$};
     \node at (29mm,19.5mm) [] {$\mf d_1$};
 \end{scope}

 \node at (138mm, 76mm) [] {CSWP};
 \node at (110mm,50mm) [anchor= east] {$\mf n, \mf m, \mathbb{C}\tb$};
 \node at (105mm,19mm) [anchor= east] {$\epsR, \kR$};
 \node at (66.5mm,48mm) [] {$\hat{\mf n} = \mf R \mf n$};
 \node at (66.5mm,43mm) [] {$\hat{\mf m} = \mf R \mf m$};
 \node at (30mm,40mm) [] {$\mf x = \textcolor{CPSgreentwo}{\mf r} + \textcolor{CPSredtwo}{ x_\alpha\mf d_\alpha}$};
 \node at (138mm, 43mm) [] {$G_h(\mf u_h, \lmlmp, \mulmp) = 0$};
\end{tikzpicture}

%% file: S2BeamTheory.tex
\section{The geometrically nonlinear beam}
\label{Sec. Beam theory}

\subsection{Kinematics}
\label{Sec. Kinematics}

In the geometrically nonlinear beam theory, a shear-deformable beam is defined as a framed curve with centerline $\mathbf{r}: s \mapsto \mathbb{R}^3$, $s\in [0,L]$ being the arc-length coordinate and $L$ the beam length, to which a local orthonormal coordinate system with directors $\mathbf{d}_i(s)$ with $i=1,2,3$ is attached. 
At each coordinate along the beam centerline, its cross-section is denoted as $\mathcal{A}(s)=\{X_\alpha \mathbf{d}_\alpha(s),\, (X_1,X_2)\in A\}$, where $A\subset\mathbb{R}^2$ would be $A=\{(X_1,X_2): |X_1|<a,|X_2|<b\}$ for a beam with rectangular cross-section or $A=\{(X_1,X_2): X_1^2+X_2^2<r^2\}$ for a circular cross-section.
The relationship between $\mathbf{d}_i$ and its reference counterpart $\mathbf{D}_{i}$ is given by the rotation  $\mathbf{R}: s \mapsto \text{SO}(3)$, so that $\mathbf{d}_i(s)=\mathbf{R}(s)\mathbf{D}_{i}(s)$. 
Without loss of generality, and in favor of a concise explanation of the theory, only initially straight beams with constant cross-sections are treated here. In that case, $\mathbf{e}_i=\mathbf{D}_i$ and hence $\mathbf{d}_i(s)=\mathbf{R}(s)\mathbf{e}_i(s)$, where $\mathbf{e}_i$ refers to the reference coordinate system. 
For initially straight beams, it holds that the global coordinate system and the referential coordinate system align, i.e., $\mf e_i = \mf E_i$. In general, the kinematic quantities that describe the beam's deformation are  $\mathbf{r}(s)$ and $\mathbf{R}(s)$. For a graphic explanation of the beam's configuration and their interplay, see \cref{fig: Beam Kinematics} (left).

The deformed configuration $\mf x \in \mathbb{R}^3 $ of the 3D beam continuum is given by
\begin{align}\label{Eq: x=r+X}
    \mathbf{x}(\mathbf{X}) &= \mathbf{r}(s)  + X_\alpha \mathbf{d}_\alpha(s) \notag \\ 
    &=  \mathbf{r}(s)  + X_\alpha \mathbf{R}(s) \mathbf{e}_\alpha  \\
    & \qquad \forall ~(X_1,X_2)\in A,~s \in [0,L], \notag
\end{align}
where $\mathbf{X} \in \mathbb{R}^3 $ refers to the coordinates $(X_1,X_2,s)$ in the reference configuration. The corresponding deformation gradient is
\begin{equation} \label{Eq: F rigid}
    \mathbf{F}(\mathbf{X}) =\mathbf{R}\big( (\epsR + [\kR]_\times X_\alpha \mathbf{e}_\alpha ) \otimes \mathbf{e}_3 + \mathbf{I} \big), 
\end{equation}
where  $\epsR(s)=(\varepsilon_{1},\varepsilon_{2},\varepsilon_{3})\T$ is the vector of the two shear strains and the axial strain, $\kR(s)=(\kappa_{1},\kappa_{2},\kappa_{3})\T$ is the vector of the two curvatures and the twist, and $[\kR]_\times$ is the skew symmetric cross-product matrix generated by $\kR$. 
Explicit dependencies on the arc-length coordinate $s$ have been omitted from \cref{Eq: F rigid} for better readability. Note that the strain measures $\epsR$ and $\kR$ uniquely describe the strain state in the beam continuum and are hence referred to in the literature as strain prescriptors \cite{herrnbock_YieldSurface_2021}. For a thorough derivation of the deformation gradient, we refer to \cite{alzate_FSB_2025,auricchio_Cosserat_2008}. 
In terms of $\mathbf{r}$ and $\mathbf{R}$, the beam strain measures are explicitly expressed as
\begin{align}
    \epsR &= \mathbf{R}^{\mathrm{T}}\mathbf{r}'-\e{3}, \label{Eq: Epsilon0}\\
    \kR &=\text{ax}\big([\kR]_\times\big) =\text{ax}\big(\mf R\T\mf R'\big). \label{Eq: Kappa0} 
\end{align}
In \cref{Eq: Kappa0}, the operator ax$(\bullet)$ extracts the axis of a skew-symmetric matrix and $\partial_s\bullet= \bullet'$ refers to the arc-length derivative. 

The deformation gradient in \cref{Eq: F rigid} can also be expressed as
\begin{equation}
    \mathbf{F}=  \mathbf{R}\, \mf F^{\{\text{cs}\}},
\end{equation}
where $\mf F^{\{\text{cs}\}} (X_\alpha;\epsR(s),\kR(s))$ can be interpreted as the stretch over the cross-section. However, note that it is not a strict polar decomposition as conventionally used in continuum mechanics, cf.\ \cite{holzapfel_nonlinear_2010}, since the strain measures $\kR, \epsR$ still depend on the rotation $\mf R$, see \cref{Eq: Epsilon0,Eq: Kappa0}. 
\subsection{Balance equations}
\label{Sec. Balance}

The equilibrium configuration of a beam is governed by the balance equations of linear and angular momenta, given by
\begin{align}
    \hatn'&= \mathbf{0}, \label{Eq: Balance LinMom} &\forall&~ s\in\,]0,L[\\
    \hatm'+ \mf r'\times \hatn &= \mathbf{0}, &\forall&~ s\in\,]0,L[\label{Eq: Balance AngMOm}
\end{align}
where $\hatn =(\hat {{n}}_1,\hat {{n}}_2,\hat {{n}}_3)\T$ denotes the shear and axial forces, and $\hatm=(\hat {{m}}_1,\hat {{m}}_2,\hat {{m}}_3)\T$ the bending and torsional moments acting on the deformed beam; see \cref{fig: Beam Kinematics}. These balance equations can be derived from the equilibrium of an infinitesimal beam segment, see \cite{alzate_FSB_2025} for further details or \cite{auricchio_Cosserat_2008, herrnbock_PhdThesis_2023} for a variational derivation.

Note that the balance equations in \cref{Eq: Balance LinMom,Eq: Balance AngMOm} are expressed in the spatial configuration. However, as will be shown throughout this work, constitutive relations are typically formulated in the reference configuration, and the corresponding quantities require a push-forward transformation. For example, the relation between spatial and referential forces and moments is given by
\begin{equation}
    \hatn = \mf R \mf n, \qquad
    \hatm = \mf R \mf m,
\end{equation}
where $\mf n= ({{n}}_1,{{n}}_2,{{n}}_3)\T$ and $\mf m=({{m}}_1,{{m}}_2,{{m}}_3)\T$ denote the forces and moments in the reference configuration, related to their spatial counterparts via the rotation matrix $\mathbf{R}$. An analogous transformation applies to the strain measures $\epsR,\kR$.

\subsection{Constitutive relations}
\label{Sec. Constitutive}

As in 3D continuum mechanics, where strains and stresses are considered to be energetic conjugate pairs associated through an energy potential, a beam internal energy density potential can also be defined in the beam theory as
\begin{equation}\label{Eq: Beam Pot. Rigid}
    \Psi^b(\epsR,\kR) = \int_A \Psi \big( \mf F(X_\alpha;\epsR,\kR)\big)~dA,
\end{equation}
where $\Psi:\mf F \mapsto \mathbb{R}$ is a hyperelastic strain energy density function defined per unit volume that depends on the deformation gradient $\mf F$ over the cross-section. 
$\Psi^b: (\epsR,\kR)\mapsto \mathbb{R}$ is designated as the beam's energy potential, defined as energy per unit length as it is integrated over the cross-section.  
Consequently, following \cite{smriti_thermoelastoplastic_2019}, the constitutive relations for the beam can be obtained in a thermodynamically consistent manner  from the beam potential as
\begin{equation} \label{Eq: Beam Pot. n and m}
    \mf n = \Psi_{,\epsR}^b, \qquad \mf m = \Psi_{,\kR}^b.
\end{equation}
Analogously, it follows for the beam's stiffness that 
\begin{equation} \label{Eq. Beam Pot. stiffness}
    \mathbb{C}^b_{\varepsilon \varepsilon} = \Psi_{,\epsR \epsR}^b , \quad
    \mathbb{C}^b_{\kappa \kappa} = \Psi_{,\kR \kR}^b, \quad  \mathbb{C}^b_{\varepsilon \kappa} = \Psi_{,\epsR \kR}^b.
\end{equation}

In linear elasticity theory, which is usually assumed in conventional beam models, the beam potential is assumed to be quadratic in the strain measures, so that
\begin{equation}\label{Eq: Beam Pot. Linear El.}
    \Psi^b =  \frac{1}{2} \epsR\T \mathbb{C}^b_{\varepsilon \varepsilon} \epsR + \frac{1}{2} \kR\T \mathbb{C}^b_{\kappa \kappa} \kR + \epsR\T \mathbb{C}^b_{\varepsilon \kappa} \kR,
\end{equation} 
with constant beam stiffness $\mathbb{C}^b_{\varepsilon \varepsilon},\, \mathbb{C}^b_{\varepsilon \kappa},\, \mathbb{C}^b_{\kappa \kappa}$.
As a consequence, the stress resultants are linear in the strain measures and the constitutive model becomes 
\begin{align} \label{Eq: Lin Const. Beam Model}
    \begin{pmatrix}
    \mf n \\
    \mf m
    \end{pmatrix}
    &=
    \begin{pmatrix}
    \mathbb{C}^b_{\varepsilon \varepsilon} & \mathbb{C}^b_{\varepsilon \kappa} \\
    {\mathbb{C}^b_{\varepsilon \kappa}}^{\mathrm{T}} & \mathbb{C}^b_{\kappa \kappa}
    \end{pmatrix}
    \begin{pmatrix}
    \epsR \\
    \kR
    \end{pmatrix} = \mathbb{C}^b \begin{pmatrix}
    \epsR \\
    \kR
    \end{pmatrix}.
\end{align}
Note that the constitutive model in \cref{Eq: Lin Const. Beam Model} is linear, does not allow cross-sectional deformation, and is limited to small elastic strains. 

%% file: S3TheCSWP.tex
\section{Cross-sectional warping problem}
\label{Sec. CSWP}

While simple and widely used, linear constitutive models for 3D beams exhibit clear limitations when large strains and nonlinear material behavior are involved. In practice, cross-sectional deformations occur, and the use of materials beyond the linear elastic regime is often of interest. However, in this context, it is not appropriate to model hyperelastic behavior in beams by directly evaluating \cref{Eq: Beam Pot. Rigid} with the deformation gradient given in \cref{Eq: F rigid}, as this leads to overly stiff responses, see \cite{schommartz_PANN_CSWP_2025}.

To account for the influence of general hyperelastic 3D material laws on beam constitutive relations, we follow the approach of Arora et al. \cite{arora_computational_2019} and formulate a cross-sectional problem that incorporates both warping and arbitrary hyperelastic material models. As introduced in \cref{Sec: Intro}, this problem is referred to as the cross-sectional warping problem (CSWP) and constitutes the central focus of this work.

\subsection{Problem formulation}
\label{Sec: Problem formulation}

Cross-sectional warping is introduced by augmenting the deformation map in \cref{Eq: x=r+X} with the warping function $\mf u:(s,X_\alpha)\mapsto\mathbb{R}^3$, defined in the reference configuration. The resulting deformation map reads
\begin{align} \label{Eq: x=r+X+u}
    \mathbf{x}(\mathbf{X}) &=  \mathbf{r}(s) + X_\alpha \R(s) \mathbf{e}_\alpha + u_i(s,X_\alpha) \R(s) \mathbf{e}_i \notag \\
    &= \mathbf{r}(s) + \mf x^{\{\text{cs}\}},
\end{align}
where $\mf x^{\{\text{cs}\}}$ denotes the deformation of the cross-section. For the remainder of this manuscript, explicit dependencies on the coordinates $\mf X$ are omitted for clarity.

Following the procedure introduced in \cref{Sec. Kinematics}, isolating the rotation $\mf R$ on the left yields
\begin{align} \label{Eq: Def. Grad. Warp.}
      \mf F&= \R\,\bigl( (\epsR + [\kR]_\times \mf x^{\{\text{cs}\}} ) \otimes \mathbf{e}_3 +\mf I+ \nabla_\alpha \mf u \bigr) \notag \\
    &= \mf R\, \mf F^{\{\text{cs}\}},
\end{align}
where $\mf F^{\{\text{cs}\}}$ corresponds to the net deformation gradient of the cross-section, now including the in-plane deformation gradient $\nabla_\alpha \mf u = \mf u_{,\alpha} \otimes \mf e_\alpha$.

In \cref{Eq: Def. Grad. Warp.}, the derivative of the warping function with respect to the arc-length coordinate $s$ is assumed to vanish, i.e., $\mf u_{,s}=0$. This assumption enforces a uniform strain field along the beam centerline and reduces the three-dimensional beam problem to its two-dimensional cross-section. Consequently, the deformation gradient $\mf F$ becomes constant along the beam axis and can, without loss of generality, be evaluated at $s=0$. At this location, the assumptions $\mf R = \mf I$ and $\mf r(s=0)=\mf 0$ can be adopted, which yield
\begin{align}
    \mf x \stackrel{!}{=}\, \mf x^{\{\text{cs}\}} &=\, X_\alpha \mathbf{e}_\alpha + u_i(s,X_\alpha)  \mathbf{e}_i \notag \\
    &=\, \mf X + \mf u, 
\end{align}
 with $\mf u:(X_1,X_2)\mapsto\mathbb{R}^3$ and consequently
\begin{align}
    \mf F\big( X_\alpha, \mf u( X_\alpha); \epsR, \kR \big) \stackrel{!}{=} \mf F^{\{\text{cs} \}}(s=0)\notag \\ 
    =\,  
    (\epsR + [\kR]_\times \mf x ) \otimes \mathbf{e}_3 
    +\mf I+ \nabla_\alpha \mf u. \label{Eq: F of CSWP}
\end{align}
In the following, the deformation gradient is consistently defined as $\mf F \stackrel{!}{=} \mf F^{\{\text{cs}\}}$.

In an analogous fashion to \cref{Eq: Beam Pot. Rigid}, the beam internal energy potential can now be expressed as
\begin{equation} \label{Eq: Beam Pot. warping}
    \Psi^b (\epsR, \kR) = \min_{\mf u} \int_A \psi\Big( \mf F \big(X_\alpha, \mf u( X_\alpha); \epsR, \kR \big) \Big)~dA.
\end{equation}
This means that for given beam strain measures $(\epsR, \kR)$, there is a unique warping function $\mf u$ that minimizes the total strain energy over the cross-section, which is then considered as the beam energy potential $\Psi^b$.
The cross-sectional warping problem thus consists of solving \cref{Eq: F of CSWP} to find $\mf u$ for given $(\epsR, \kR)$.
To guarantee an overall unique solution, rigid body motions need to be constrained. For this, as in \cite{arora_computational_2019,herrnbock_YieldSurface_2021}, we use a Lagrange multiplier approach.

\subsection{Lagrange multipliers}

In order to fix the cross-section in space, the following conditions need to be satisfied
\begin{equation} \label{Eq: Fix CS conditions}
    \int_A \mf x ~dA = \mathbf{0}, \qquad
    \int_A \mx ~dA = \mathbf{0},
\end{equation}
where $\mx$ is defined as 
\begin{equation}
    \mx
=
\begin{pmatrix}
x_2  x_3 \\
x_1 x_3 \\
\text{arctan}(\frac{x_2}{x_1})-\text{arctan}(\frac{X_2}{X_1})
\end{pmatrix}.
\end{equation}
While \cref{Eq: Fix CS conditions}$_1$ fixes the cross-section to its center of mass, \cref{Eq: Fix CS conditions}$_2$ avoids rotation around its main axes. These conditions are imposed weakly through the Lagrange multipliers $\lmlmp \in \mathbb{R}^3$ and $\mulmp \in \mathbb{R}^3$ and the penalty term
\begin{equation} \label{Eq: Lagrange Multiplers Pot.}
    \mathcal{L}(\mf u,\lmlmp,\mulmp) = \int_A \lmlmp \cdot \mf x +\mulmp \cdot  \mx ~dA.
\end{equation}
The following treatment of the Lagrange multiplier method is analogous to the one presented in \cite{herrnbock_PhdThesis_2023}.

\subsection{Constrained minimization problem}

Combining  \cref{Eq: Beam Pot. warping,Eq: Lagrange Multiplers Pot.}, the CSWP can be formulated as a constrained minimization problem for the following functional:
\begin{equation}
    \mathcal{F}(\mf u,\lmlmp,\mulmp) =\int_A \psi(\mf F) + \mathcal{L}(\mf u,\lmlmp,\mulmp) ~dA.
\end{equation}
The minimization problem can thus be formulated as: find $\mf u,\mulmp, \lmlmp$ so that the functional $\mathcal{F}$ is minimized. 

A stationary solution is obtained by setting the first variation of the functional to zero as
\begin{equation} \label{Eq: G}
    G = \delta \mathcal{F} =\int_A \delta \psi + \delta \mathcal{L}~ dA \stackrel{!}{=}0,
\end{equation}
where explicit dependencies have been omitted here for better readability. The terms of this variational, weak form of the CSWP can be expanded into 
\begin{align}
    \int_A \delta \psi~dA =& \int_A \mf S:\delta \mf E~dA \label{Eq: WeakForm psi} ,\\
    \int_A \delta \mathcal{L} ~dA =& \int_A \mf x \cdot \delta \lmlmp  + \lmlmp \cdot \delta \mf u+ \mx \cdot \delta  \mulmp \notag \\
    &~~ + (\mathbf{\mathfrak{M}} \mulmp) \cdot \delta \mf u ~dA, \label{Eq: WeakForm LMP}
\end{align}
with 
\begin{equation}
    \delta \mx = \mathbf{\mathfrak{M}}\T \cdot \delta \mf u =
    \begin{pmatrix}
0 & x_3 & x_2 \\
x_3 & 0 &x_1 \\
-\frac{x_2}{x_1^2+x_2^2} & \frac{x_1}{x_1^2+x_2^2} & 0
\end{pmatrix}
\delta \mf u.
\end{equation}

The problem stated in \cref{Eq: G} is nonlinear and its iterative numerical solution using Newton's method requires its linearization
\begin{equation}
\Delta G = \int_A \Delta \delta\psi + \Delta \delta \mathcal{L} ~dA. 
\label{Eq: DeltaG}
\end{equation}
Here, it is
\begin{align}
    \Delta \delta \psi =& \Delta \mf S : \delta \mf E + \mf S : \Delta \delta \mf E \label{Eq: Lin delta psi} ,\\
    \Delta  \delta \mathcal{L} =& \Delta \mf u \cdot \delta \lmlmp +\Delta \lmlmp \cdot \delta \mf u  +\mathfrak{M}\T \cdot \Delta \mf u \cdot \delta \mulmp \notag \\
    &+\mf \Xi \cdot \Delta \mf u \cdot \delta \mf u  + \mathfrak{M} \cdot \Delta \mulmp \cdot \delta \mf u ,  \label{Eq: Lin delta L}
\end{align}
with 
\begin{align}
        \mathbf{\Xi} =
    \begin{pmatrix}
\XiI & \XiII & \mu_2 \\
\XiII & -\XiI &\mu_1 \\
\mu_2 & \mu_1 & 0
\end{pmatrix}
\end{align}
and
\begin{align}
    \delta \mf E &= \frac{1}{2} \left( \mf F \T \delta \mf F + \delta  \mf F \T  \mf F \right), \label{Eq: delta E} \\
    \delta \mf F &=  \Bigl(\delta \mf u_{,1},~\delta \mf u_{,2}, ~[\kR]_\times \delta \mf u\Bigr).\label{Eq: delta F}
\end{align}

In summary, to determine the beam internal energy potential $\Psi^b$, the variational problem $G=0$ stated in \cref{Eq: G}, expressed here in a fully material formulation using the PK2 stress and GL strain tensors, needs to be solved for the warping function $\mf u$ and the Lagrange multipliers $\lmlmp, \mulmp$, using its linearization $\Delta G$ provided in \cref{Eq: DeltaG}. 

\subsection{Stress resultants and beam stiffness} 
\label{Sec: Str. Res. and Stiff.}

After solving the CSWP and determining the warping function $\mf u$ for given $(\epsR,\kR)$, the stress resultants, i.e., the resultant forces $\mf n$ and moments $\mf m$ of the beam theory can be obtained by integrating the traction $\mf T$ over the reference cross-section:
\begin{align}
    \mf n & = \int_A \mf T ~dA  \stackrel{!}{=}  \Psi_{,\epsR}^b, \label{Eq: n0 integral form} \\
    \mf m & = \int_A \mf x \times  \mf T ~dA \stackrel{!}{=} \Psi_{,\kR}^b. \label{Eq: m0 integral form}
\end{align}
The traction $\mf T = \mf P \mf e_3$ is defined as the normal projection of the first Piola--Kirchoff stress tensor $\mf P$ onto the reference cross-section, where the first and second Piola-Kirchoff stress tensors are related through $\mf P = \mf F \mf S$. 
The equivalence between the stress resultants and the derivatives of the implicitly defined potential $\Psi^b$ is further detailed in  \cite{herrnbock_PhdThesis_2023, arora_computational_2019}.
Alternatively, \cref{Eq: n0 integral form,Eq: m0 integral form} can be written in a compact manner as
\begin{equation}
    \Psi_{,p}^b = \int_A \big( \mf P \mf e_3 \big) \cdot \big( \epsR_{,p} +[\kR_{,p}]_\times \mf x \big)~dA,
\end{equation}
where $p \in \{\varepsilon_{1},\varepsilon_{2},\varepsilon_{3}, \kappa_{1}, \kappa_{2}, \kappa_{3} \}$ corresponds to one of the six beam strain measures and thus, e.g., $\epsR_{,\varepsilon_1}=(1,0,0)\T$. 
Note that here and in the following, partial derivatives w.r.t.\ the Cartesian coordinates $X_1$ and $X_2$ are denoted by $\bullet_{,1}$ and $\bullet_{,2}$, whereas derivatives with respect to the strain measures are denoted by $\bullet_{,p}$ or $ \bullet_{,q}$.

Using the same notation, the coefficients of the beam stiffness matrix $\mathbb{C}^b\in \mathbb{R}^{6 \times6}$ from \cref{Eq. Beam Pot. stiffness} are obtained as, see also \cite{herrnbock_YieldSurface_2021, arora_computational_2019}:
\begin{align}\label{Eq:NL Const. Beam Model}
    \mathbb{C}_{pq}^b =& \int_A \big(\mathbb{A}:\mf F_{,q} \big)\mf e_3 \cdot \big(\epsR_{,p} + [\kR_{,p}]_\times \mf x \big) \notag \\
    & + \big(\mf P \mf e_3\big)\cdot \big([\kR_{,p}]_\times \mf u_{,q}\big)~dA \notag \\
    \stackrel{!}{=}&\, \Psi_{,pq}^b
\end{align}
Here, $\mathbb{A}=\frac{\partial \mf P}{\partial \mf F}=\frac{\partial^2\psi}{\partial\mf F^2}$ corresponds to the tangent stiffness of the constitutive model of the material. 
Furthermore, the expression $\mf F_{,q}$ can be expressed explicitly in a column-wise fashion as
\begin{align}
    \mf F_{,q} =&~  \Bigl( \mf u_{,q1},~ \mf u_{,q2},~  \epsR_{,q} + [\kR_{,q}]_\times \mf x +[\kR]_\times \mf u_{,q} \Bigr).
\end{align}
The actual determination of $\mf u_{,q}$, the sensitivity of the warping function $\mf u$ w.r.t.\ the strain measure $q$, and its material derivatives $\mf u_{,q\alpha}$, is non-trivial and will be discussed in \cref{Sec: Calc. Beam Stiff.}.

%% file: S4IGA.tex
\section{Isogeometric finite element discretization}
\label{Sec. IGA}

Based on the fully material formulation of the hyperelastic CSWP in terms of the Green-Lagrange strain and second Piola-Kirchhoff stress tensors introduced in \cref{Eq: G,Eq: DeltaG}, an efficient isogeometric finite element implementation using the Voigt notation is presented in this section. In addition to determining the warping function $\mf u$, the proposed computational framework enables the efficient evaluation of the stress resultants $\mf n,\mf m$, the sensitivities $\mf u_{,q}$, and consequently also the beam stiffness $\mathbb{C}^b$. 
The discretization employs NURBS or B-spline basis functions and builds upon the isogeometric finite element implementation provided by the open-source library \texttt{NLIGA} \cite{du_nliga_2020}. 
It is emphasized, however, that the proposed formulation of the CSWP in Voigt notation is not restricted to isogeometric analysis and can also be easily implemented within a classical finite element framework.

\subsection{Discretization of the CSWP} \label{Sec: Discretization}

The discretization of the two-dimensional cross-section and the associated warping function is based on $n$ bivariate B-spline basis functions $N_I(\xi,\eta):\hat\Omega\to\mathbb{R}$ defined over a parametric domain $\hat\Omega\subset\mathbb{R}^2$. 
For details on the definition of B-spline functions, see for instance \cite{piegl2012nurbs}. Following the isoparametric concept, the discretized initial and deformed configurations of the cross-section, as well as the warping displacement, are then approximated as
\begin{align}
    \mf X &\approx \mf X_h(\xi,\eta) = \sum_{I=1}^{n} N_I(\xi,\eta)\, \mf X_I, \label{Eq. X Discr}   \\
    \mf x &\approx \mf x_h(\xi,\eta) \;= \mf X_h(\xi,\eta) + \mf u_h(\xi,\eta), \\
    \mf u &\approx \mf u_h(\xi,\eta) \;= \sum_{I=1}^{n} N_I(\xi,\eta)\, \mf u_I, \label{Eq. Discr}  
\end{align}
where $\mf X_I \in \mathbb{R}^3$ and $\mf u_I\in \mathbb{R}^3$ denote the control points associated with the $n$ basis functions. 
Although the CSWP is defined over a two-dimensional manifold, $\mf X:\hat\Omega\mapsto\mathcal{A}$, each control point possesses three degrees of freedom, thereby allowing for both in-plane and out-of-plane cross-sectional deformations.

Following a standard Galerkin procedure, the corresponding variations of the displacement field and the deformation gradient, see \cref{Eq: delta F}, are approximated as
\begin{align}
    \delta \mf u &\approx \delta \mf u_h(\xi,\eta) = \sum_{I=1}^{n} N_I(\xi,\eta)\, \delta \mf u_I, \label{Eq: Disctr. delta u} \\
    \delta \mf F &\approx \delta \mf F_h(\xi,\eta) = \sum_{I=1}^{n} \delta \mf F_I \notag\\
    &= \sum_{I=1}^{n} \Bigl(N_{I,1}\delta \mf u_I,~N_{I,2}\delta \mf u_I, ~N_I[\kR]_\times \delta \mf u_I \Bigr), \label{Eq: Disctr. F}
\end{align}
where the material derivatives are obtained via the chain rule and the Jacobian of the geometric discretization \cref{Eq. X Discr}   as
\begin{align}
    \nabla N_{I} = \begin{pmatrix}
        \partial_1 N_{I} \\ \partial_2 N_{I}
    \end{pmatrix}\T  =  \begin{pmatrix}
        \partial_{\xi} N_{I} \\ \partial_{\eta} N_{I}
    \end{pmatrix}\T \begin{pmatrix}
        \partial_{\xi} X_1 & \partial_{\eta} X_1 \\ 
        \partial_{\xi} X_2 & \partial_{\eta} X_2
    \end{pmatrix}^{-1}.
\end{align}

\paragraph{Discretized variational form $G_h$.}
These discretizations are now substituted into the variational form given by \cref{Eq: G}:
\begin{equation}
    G_h  = G(\mf u_h, \lmlmp,\mulmp) =  \int_A \delta \psi_h + \delta \mathcal{L}_h~dA \stackrel{!}{=}0. \label{Eq: Disctr. G}
\end{equation}
According to \cref{Eq: WeakForm psi,Eq: WeakForm LMP}, the individual contributions can be expressed as
\begin{align}
    \int_A \delta \psi_h ~ dA \;=& \int_A \sum_{I=1}^{n} \mf S : \delta \mf E_I ~dA, \label{Eq: delta psi_h} \\
    \int_A \delta \mathcal{L}_h ~dA \; =& \int_A \mf x_h \cdot \delta \lmlmp  +  \mxh \cdot \delta  \mulmp \notag \\
    &+ (\mathbf{\mathfrak{M}} \mulmp + \lmlmp ) \cdot \sum_{I=1}^{n} N_I \delta \mf u_I ~dA, \label{Eq: delta L_h}
\end{align}
where
\begin{align}
    \delta \mf E_I =& \frac{1}{2} \big( \mf F \T \delta \mf F_I + \delta  \mf F_I \T  \mf F \big) = \big( \mf F \T \delta \mf F_I \big)^{(s)} \notag \\
    =&  \Bigl(N_{I,1}\mf F\T \delta \mf u_I, \,
    N_{I,2}\mf F\T \delta \mf u_I, \notag \\
    &~~ N_I\mf F\T[\kR]_\times \delta \mf u_I\Bigr)^{(s)}. \label{Eq: delta E_I}
\end{align}
The superscript $(\bullet)^{(s)}=\tfrac{1}{2}(\bullet+\bullet\T)$ denotes the symmetric part of $\delta \mf E_I$. For readability, the discretization index $(\bullet)_h$ is omitted in the following.

Owing to the symmetry of $\mf S$ and $\delta \mf E_I$, the integrand in \cref{Eq: delta psi_h} can be rewritten in vector form as
\begin{align}
    \sum_{I=1}^{n} \mf S : \delta \mf E_I = \sum_{I=1}^{n} \delta \underline{\mf E}\T_I \underline{\mf S},
\end{align}
where the Voigt notation is adopted as
\begin{align}
    \delta \underline{\mf E}_I\T &= \bigl(\delta E_{11}, \delta E_{22}, \delta E_{33},  2\delta E_{12},2 \delta E_{23},2\delta  E_{13} \bigl)_I \notag \\
    &= \bigl(\delta E_{1},~ \delta E_{2},~ E_{3},~\delta  E_{4},~\delta E_{5},~\delta E_{6} \bigl)_I, \\
    \underline{\mf S}\T &= 
    (S_{11},~S_{22}, S_{33},~ S_{12},~ S_{23},~ S_{13}) \notag \\ 
    &=( S_{1},~ S_{2}, ~S_{3},~ S_4,~ S_5,~ S_6).
\end{align}
Note that this numbering is consistent with \cite{wriggers_NLFEM_2008} and the implementation in \texttt{NLIGA} \cite{du_nliga_2020}, although the off-diagonal coefficients are ordered differently in the classical Voigt notation.

Since $\delta \mf E_I$ is linear in $\delta \mf u_I$, the relation can be written as
\begin{equation}
    \delta \underline{\mf E}_I\T =\big(\mf B_I \delta \mf u_I \big)\T = \delta \mf u_I\T \mf B\T_I, \label{Eq: delta E_I^T}
\end{equation}
where $\mf B_I\T \in \mathbb{R}^{3 \times 6}$ is the strain--displacement operator, which is obtained as
\begin{align}
     \mf B_I\T =& \Bigl(
     \underline{\mf F}_1 N_{I,1}, \;
      \underline{\mf F}_2 N_{I,2}, \;
      -N_I [\kR]_\times \underline{\mf F}_3, \notag \\
     &~~ \underline{\mf F}_1 N_{I,2}+\underline{\mf F}_2 N_{I,1},\;
     \underline{\mf F}_3 N_{I,2}-N_I [\kR]_\times \underline{\mf F}_2,\notag \\
     &~~\underline{\mf F}_3 N_{I,1}-N_I [\kR]_\times \underline{\mf F}_1 \Bigr), \label{Eq: B_I^T}
\end{align}
with the column vectors $\underline{\mf F}_i\in\mathbb{R}^3, i=1,2,3$, such that $\mf F = (\underline{\mf F}_1, \underline{\mf F}_2, \underline{\mf F}_3)$.

Since $G_h=0$ must hold for any $\delta \mf u,\, \delta \lmlmp,\, \delta \mulmp$  in a minimum $(\mf u_h,\lmlmp,\mulmp)$, and thus for any $\delta\mf u_I$, the following residual vectors must all be zero:
\begin{align}
    \mf f^{u}_I &= \int_A \mf B_I \T \underline{\mf S} + \mf (\lmlmp + \bm {\mathfrak{M}} \cdot \mulmp)N_I~dA &\stackrel{!}{=} \mf 0 ,\label{Eq: f^u}\\
    {\mf f}^\lambda &= \int_A \mf x_h ~dA &\stackrel{!}{=} \mf 0,  \label{Eq: f lambda}\\
    {\mf f}^\mu &= \int_A \mxh ~dA &\stackrel{!}{=} \mf 0. \label{Eq: f mu} 
\end{align}
Note here that $\mf B_I,\, \underline{\mf S},\, \bm {\mathfrak{M}}$, and $\mf x_h$ all depend on $\mf u_h$ and its coefficients $\mf u_J$.

\paragraph{Solution scheme.}
\Cref{Eq: f^u,Eq: f lambda,Eq: f mu} form a $3(n+2)$-dimensional nonlinear system of equations for the control point displacement vector $\mf U=(\mf u_1, \ldots, \mf u_n)$, together with $\lmlmp \in \mathbb{R}^3$ and $\mulmp\in \mathbb{R}^3$:
\begin{align}
\hat{\mf F}(\hat{\mf U})&= \mf 0, \label{Eq: Fhat}\\
 \text{with}\quad\hat{\mf U} &= (\mf U, \lmlmp, \mulmp)\T, \\
    \hat{\mf F} &= ({\mf f}^u, {\mf f}^\lambda, {\mf f}^\mu)\T.
\end{align}

Newton's method is employed to iteratively solve this discretized CSWP, requiring the solution of the linear system
\begin{equation}
    \hat{\mf K}(\hat{\mf U}_k) \Delta \hat{\mf U}_k = -\hat{\mf F}(\hat{\mf U}_k), \label{Eq: KU=-F}
\end{equation}
in each iteration $k$.
Here,
\begin{align}
    \hat{\mf K}(\hat{\mf U}) &= \begin{pmatrix}
        {\mf K^{uu}} & {\mf K^{u \lambda}} & {\mf K^{u \mu}} \\
        {\mf K^{\lambda u}} & \mf 0 & \mf 0 \\
        {\mf K^{\mu u}} & \mf 0 & \mf 0 \label{Eq: Khat}
        \end{pmatrix},
\end{align}
is the tangent stiffness matrix, which will be derived below.
Note that $\hat{\mf K}$ is a symmetric matrix, as $\mf K^{uu}=\mf K^{uu \T}$, $\mf K^{\lambda u}=\mf K^{u \lambda \T}$, $\mf K^{\mu u}=\mf K^{u \mu \T}$, and sparse, as the B-spline basis functions $N_I$ generally have limited, local support.

The solution of \cref{Eq: KU=-F} is followed by the update step
\begin{equation}
    \hat{\mf U}_{k+1} = \hat{\mf U}_k + \Delta \hat{\mf U}_k.
\end{equation}
For incrementally increasing prescribed $(\epsR,\kR)$, the solution at each load step is considered converged once the iteration error falls below a prescribed tolerance.

\paragraph{Tangent stiffness matrix.}
For solving the nonlinear system of equations \cref{Eq: Fhat}, the tangent stiffness matrix is required. Relating to the linearization $\Delta G$ in \cref{Eq: DeltaG}, the $3\times3$ sub-matrices are obtained as:
\begin{align}
    \mf K_{IJ}^{uu} =& ~\frac{\partial \mf f^{u}_I}{\partial \mf u_J} 
    = \int_A \mf B_I\T\frac{\partial \underline{\mf S}}{\partial \underline{\mf E}} \mf B_J  + \mathfrak{B}_{IJ} \bigl( \underline{\mf S} \otimes \mf I\bigr)  \notag \\
    &\qquad\qquad\quad + N_I \mf \Xi N_J ~dA, \label{Eq: K u u}\\
    \mf K^{u\lambda}_I =&~ \frac{\partial \mf f^{u}_I}{\partial \lmlmp} =\int_A N_I \mf I ~dA, \label{Eq: K u lambda}\\
    \mf K^{u \mu}_I =&~ \frac{\partial \mf f^{u}_I}{\partial \mulmp} = \int_A N_I \bm {\mathfrak{M}}~dA, \label{Eq: K u mu}\\
    \mf K^{\lambda u}_J =&~ \frac{\partial \mf f^{\lambda}}{\partial \mf u_J} =\int_A N_J \mf I ~dA, \label{Eq: K lambda u}\\
    \mf K^{\mu u}_J =&~ \frac{\partial \mf f^{\mu}}{\partial \mf u_J} = \int_A N_J \bm {\mathfrak{M}}\T~dA. \label{Eq: K mu u}
\end{align}
Here, $\mathfrak{B}_{IJ} \in \mathbb{R}^{3 \times 18}$, appearing in the second term of \cref{Eq: K u u}, corresponds to a to-be-defined operator and $\underline{\mf S} \otimes \mf I\in\mathbb{R}^{18\times 3}$ denotes the Kronecker product between $\underline{\mf S}$ and the identity tensor $\mf I \in \mathbb{R}^{3\times 3}$. 

To derive the $\mathfrak{B}_{IJ}$-operator, we first introduce a column-wise representation of the strain--displacement operator $\mf B_I \T \in \mathbb{R}^{3 \times 6}$ as
\begin{align}
    \mf B_I\T &= (\underline{\mf B}_{1I},\underline{\mf B}_{2I},\underline{\mf B}_{3I},\underline{\mf B}_{4I},\underline{\mf B}_{5I},\underline{\mf B}_{6I}), \label{Eq: B_I^T symbol}
\end{align}
where the individual components $\underline{\mf B}_{iI} \in \mathbb{R}^3$, $i=1,\dots,6$, are directly obtained from \cref{Eq: B_I^T}. 
Now, the gradient of $\underline{\mf B}_{iI}$ w.r.t.\ the displacement control point $\mf u_J$ can be expressed as
\begin{align}
    \frac{\partial \underline{\mf B}_{iI} }{\partial \mf u_J}  &= \tilde{\mathfrak{B}}_{iIJ} \in \mathbb{R}^{3\times3}. \label{Eq: BiIJ}
\end{align}

As follows from \cref{Eq: B_I^T}, the operator $\mf B_I$ depends linearly on $\mf F$, which itself is linear in $\mf u_h$. Consequently, the computation of $\tilde{\mathfrak{B}}_{iIJ}$ requires the derivatives:
\begin{align}
    \frac{\partial \underline{\mf F}_1}{\partial \mf u_J} &=N_{J,1}\mf I, \notag \\
    \frac{\partial \underline{\mf F}_2}{\partial \mf u_J} &=N_{J,2}\mf I, \label{Eq: Lin of F_i} \\
    \quad \frac{\partial \underline{\mf F}_3}{\partial \mf u_J} &=N_J[\kR]_\times \mf I. \notag
\end{align} 
After substituting \cref{Eq: Lin of F_i} into \cref{Eq: BiIJ} and using \cref{Eq: B_I^T}, the following expressions are obtained:
\begin{equation}
\begin{aligned} \label{Eq: B_IJ 1-4}
    \tilde{\mathfrak{B}}_{1IJ} &=\frac{\partial \underline{\mf B}_{1I}}{\partial \mf u_J} = N_{J,1}N_{I,1} \mf I,  \\
    \tilde{\mathfrak{B}}_{2IJ} &= \frac{\partial \underline{\mf B}_{2I}}{\partial \mf u_J} = N_{J,2}N_{I,2} \mf I,  \\
    \tilde{\mathfrak{B}}_{3IJ} &= \frac{\partial \underline{\mf B}_{3I}}{\partial \mf u_J} = -N_{J}N_{I} [\kR]_\times^2,  \\
    \tilde{\mathfrak{B}}_{4IJ} &= \frac{\partial \underline{\mf B}_{4I}}{\partial \mf u_J} = (N_{J,2}N_{I,1} + N_{J,1}N_{I,2})\mf I,  \\
    \tilde{\mathfrak{B}}_{5IJ} &= \frac{\partial \underline{\mf B}_{5I}}{\partial \mf u_J} = (N_{I,2}N_J- N_{J,2}N_I)[\kR]_\times,  \\
    \tilde{\mathfrak{B}}_{6IJ} &= \frac{\partial \underline{\mf B}_{6I}}{\partial \mf u_J} =  (N_{I,1}N_J- N_{J,1}N_I)[\kR]_\times.
\end{aligned} 
\end{equation}
From this, the second term in the integral in \cref{Eq: K u u} can be expressed as
\begin{equation}
    \sum_{i=1}^6 \frac{\partial \underline{\mf B}_{iI} }{\partial \mf u_J} S_i 
    = \sum_{i=1}^6 \tilde{\mathfrak{B}}_{iIJ} S_i 
    = \mathfrak{B}_{IJ} \bigl( \underline{\mf S} \otimes \mf I\bigr),
\end{equation}
where the components $\tilde{\mathfrak{B}}_{iIJ}$ are assembled into the matrix $\mathfrak{B}_{IJ}\in \mathbb{R}^{3\times 18}$ as
\begin{align}
    \mathfrak{B}_{IJ} = \bigl(\tilde{\mathfrak{B}}_{1IJ}, \tilde{\mathfrak{B}}_{2IJ},\tilde{\mathfrak{B}}_{3IJ},\tilde{\mathfrak{B}}_{4IJ},\tilde{\mathfrak{B}}_{5IJ},\tilde{\mathfrak{B}}_{6IJ} \bigr).  \label{Eq: BIJ transposed} 
\end{align}

\subsection{Calculation of the beam stiffness $\mathbb{C}^b$}
\label{Sec: Calc. Beam Stiff.}

As discussed in \cref{Sec: Str. Res. and Stiff.}, the sensitivities $\mf u_{,q}$ of the warping function $\mf u$ with respect to the strain measures $q\in\{\varepsilon_1,\varepsilon_2, \varepsilon_3, \kappa_1, \kappa_2, \kappa_3  \}$ are required for the evaluation of the coefficients $\mathbb{C}^b_{pq}$ of the beam stiffness matrix $\mathbb{C}^b \in \mathbb{R}^{6 \times 6}$, see \cref{Eq:NL Const. Beam Model}. However, these sensitivities are not directly available from the solution vector $\hat{\mf U}$ obtained from solving \cref{Eq: Fhat}.

Typically, deriving the sensitivities $\mf u_{,q}$ is accompanied by the formulation of an additional weak form and its finite element discretization, which involves complex linearizations and extensive derivations, see \cite{arora_computational_2019,singh_CSWP_SurfaceEnergy_2022, herrnbock_PhdThesis_2023,kumar_growth-CSWP_2026}. Here, we circumvent this issue by introducing an adjoint equation derived directly from the discretized CSWP.

\paragraph{Derivation of adjoint equation.}
To compute the sensitivities, we proceed as in PDE-constrained nonlinear optimization, where the adjoint method is commonly used to obtain the gradient of a function w.r.t.\ its implicit parameters \cite{OptimizationPDEConstraints2009}.

So far, we have regarded the strain measures $q$ as fixed parameters in the formulation of the variational form of the CSWP $G$ and its discretized counterpart $G_h$, see \cref{Eq: G,Eq: Disctr. G}, from which, the nonlinear system of equations in \cref{Eq: Fhat} is derived.
Now, if we consider $q$ as variable, the displacement $\mf u_h$ and its control points $\mf u_I$, as well as the Lagrangian multipliers $\lmlmp,\mulmp$ are in fact $q$-dependent through the implicit relations
\begin{align}
    G_h(\mf u_h(q), \lmlmp(q),\mulmp(q)) &= 0, \\
    \Rightarrow\quad
    \hat{\mf F}\big(\hat{\mf U}(q)\big) &=\bf 0 \label{Eq: hatFq}.
\end{align}
Directly continuing with the fully discretized nonlinear system of equations, the residual vector $\hat{\mf F}$ must be stationary w.r.t.\ $q$ as $\hat{\mf U}(q)$ is always determined such that \cref{Eq: hatFq} is satisfied. Hence, it follows that
\begin{align}
    \frac{d \hat{\mf F}}{d q} &\stackrel{!}{=} \mf 0, \notag \\
    \Leftrightarrow \quad \frac{\partial \hat{\mf F}}{\partial \hat{\mf U}}\cdot \frac{\partial \hat{\mf U}}{\partial q} +\frac{\partial \hat{\mf F}}{\partial q} &= \mf 0, \label{Eq: F_q = K Y} \\
    \Leftrightarrow \quad \hat{\mf K} \hat{\mf Y} &= - \hat{\mf F}_{,q}. \notag
\end{align}
where $\hat{\mf K}$ denotes the tangent stiffness matrix from \cref{Eq: Khat}. 
The unknowns of the linear system in \cref{Eq: F_q = K Y} are $\hat{\mf Y}=(\mf U_{,q},\lmlmp_{,q},\mulmp_{,q})$, which comprise the vector of displacement control point sensitivities $\mf U_{,q}=(\mf u_{1,q},...,\mf u_{n,q})$ as well as the sensitivities of the Lagrange multipliers with respect to $q$, i.e., $\lmlmp_{,q}$ and $\mulmp_{,q}$.
$\hat{\mf F}_{,q}$ describes the sensitivities of the residual vector, which are derived below.

After solving the linear system of \cref{Eq: F_q = K Y}, for which $\hat{\mf K}$ can be used directly from the last Newton iteration with the converged $\hat{\mf U}$,  the sensitivity $\mf u_{,q} $ and its derivatives with respect to the coordinates $X_\alpha$ can be obtained directly from the finite element interpolation of \cref{Eq. Discr}:
\begin{align}
   \mf u_{,q} =  \sum_{I=1}^n N_{I}\mf u_{I,q}, \qquad
   \mf u_{,q\alpha} =  \sum_{I=1}^n N_{I,\alpha}\mf u_{I,q}. \label{Eq: u_q and u_q1}
\end{align}
In total, the computation of the beam stiffness $\mathbb{C}_{pq}^b$ requires six sensitivities, each corresponding to one strain measure $q\in\{\varepsilon_1,\varepsilon_2, \varepsilon_3, \kappa_1, \kappa_2, \kappa_3  \}$, see \cref{Eq:NL Const. Beam Model}. The overall procedure is summarized in \cref{algo1}.

\begin{algorithm}
\caption{Beam stiffness matrix $\mathbb{C}^b$}
\label{algo1}
\begin{algorithmic}[1]
\Require $\hat{\mathbf{K}}$ from \cref{Eq: KU=-F}
\State $q = (\varepsilon_1,\varepsilon_2, \varepsilon_3, \kappa_1, \kappa_2, \kappa_3 )\T$
\For{$j=1$ to $6$}
    \State Evaluate $\hat{\mathbf{F}}_{,q}(q_j)$ from \cref{Eq: fy I,Eq: fy lambda,Eq: fy mu}
    \State
        Solve $\hat{\mathbf{K}}\hat {\mathbf{Y}} = -\hat{\mathbf{F}}_{,q}$
    \State $\hat{\mf {U}}_{q} \gets \hat {\mathbf{Y}} $
    \For{$i=1$ to $6$}
        \State $\mathbb{C}^b_{ij}~$ from \cref{Eq:NL Const. Beam Model} using \cref{Eq: u_q and u_q1}
    \EndFor
\EndFor
\end{algorithmic}
\end{algorithm}

\paragraph{Sensitivities of the residual vector.}
For solving \cref{Eq: F_q = K Y}, the derivation and evaluation of the sensitivities of the residual vector $\hat{\mf F}_{,q}$ is required. Accordingly, taking the partial derivative of \cref{Eq: f^u,Eq: f lambda,Eq: f mu} with respect to the strain measure $q$ yields
\begin{align} 
    \mf f^{u}_{I,q} =&~ \int_A \mf B_I\T 2 \frac{\partial \underline{\mf S}}{\partial \underline{\mf C}}\underline{\mf E}_{,q} +  {\mf B_{I,q}\T}\  \underline{\mf S} ~dA   \label{Eq: fy I}\\
    \mf f^{\lambda}_{,q} =&~ \int_A \mf 0 ~dA, \label{Eq: fy lambda}\\
    \mf f^{\mu}_{,q} =&~ \int_A  \mf 0 ~ dA \label{Eq: fy mu}
\end{align}

To derive $\underline{\mf E}_{,q}$, appearing in the first term of \cref{Eq: fy I}, we first consider its tensorial form
\begin{align}
    \mf E_{,q} =  \frac{1}{2} \Bigl( \mf F_{,q}\T \mf F + \mf F\T \mf F_{,q}\Bigr) =  \Bigl( \mf F\T \mf F_{,q} \Bigr)^{(s)}, \label{Eq: E_q tensorial}
\end{align}
where it follows from \cref{Eq: F of CSWP} that
\begin{align}
    \mf F_{,q} &=  \Bigl( \mf 0,~ \mf 0,~  \epsR_{,q} + [\kR_{,q}]_\times \mf x \Bigr). \label{Eq: d_pF}
\end{align}
In Voigt notation, \cref{Eq: E_q tensorial} can be expressed as
\begin{equation}
    \underline{\mf E}_{,q} = \Bigl( 0,~  0,~E_{3,q},~  0,~E_{5,q},~E_{6,q} \Bigr)\T, \label{Eq: Eq at y0}
\end{equation}
where
\begin{align}
    E_{3,q}  &=\underline{\mf F}_3 \T (\mf \epsR_{,q} + [\kR_{,q}]_\times \mf x ), \notag  \\ 
    E_{5,q}  &=\underline{\mf F}_2 \T (\mf \epsR_{,q} + [\kR_{,q}]_\times \mf x ),  \\
    E_{6,q}  &=\underline{\mf F}_1 \T (\mf \epsR_{,q} + [\kR_{,q}]_\times \mf x ).\notag
\end{align}

Furthermore, to determine ${\mf B_{I,q}\T}$, appearing in the second term of \cref{Eq: fy I}, we recall the definition of the $\mf B_I\T$-operator in \cref{Eq: B_I^T}. Its partial derivative w.r.t the strain measure $q$ leads to
\begin{align}
    {\mf B_{I,q}\T}&   = \Bigl( \underline{\mf B}_{1I,q},\underline{\mf B}_{2I,q},\underline{\mf B}_{3I,q},\underline{\mf B}_{4I,q},\underline{\mf B}_{5I,q},\underline{\mf B}_{6I,q} \Bigr) \notag \\
    &=  \Bigl( \mf 0,~ \mf 0,\underline{\mf B}_{3I,q},~ \mf 0,\underline{\mf B}_{5I,q},\underline{\mf B}_{6I,q} \Bigr) , \label{Eq: BIpT}
\end{align}
where
\begin{align}
     \underline{\mf B}_{3I,q} &= -N_I  \Bigl( [\kR_{,q}]_\times \underline{\mf F}_3 + [\kR]_\times (\epsR_{,q} + [\kR_{,q}]_\times \mf x)  \Bigr), \notag \\
     \underline{\mf B}_{5I,q}&=  N_{I,2} (\epsR_{,q} + [\kR_{,q}]_\times \mf x)- N_I [\kR_{,q}]_\times \underline{\mf F}_2, \\ 
     \underline{\mf B}_{6I,q} &= N_{I,1} (\epsR_{,q} + [\kR_{,q}]_\times \mf x)- N_I [\kR_{,q}]_\times \underline{\mf F}_1. \notag
\end{align}

%% file: S5Results.tex
\section{Results}
\label{Sec. Results}

\begin{figure*}[t!]
    \centering
    \begin{subfigure}{0.48\textwidth}
        \centering
        \includegraphics[width=0.5\textwidth]{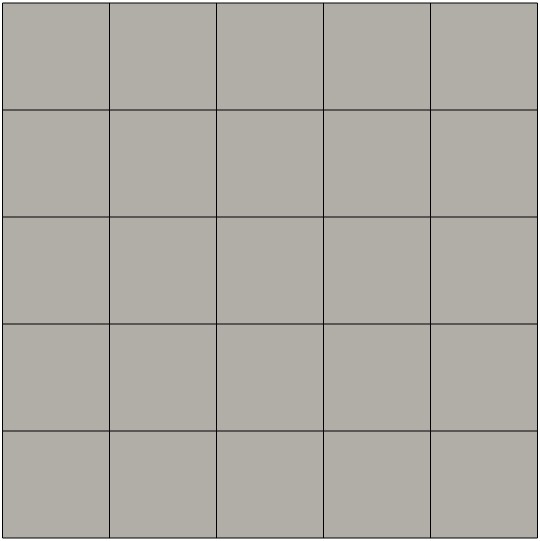}
        \caption{Isogeometric mesh of unit square cross-section.}
        \label{fig:mesh_square}
    \end{subfigure}
    \begin{subfigure}{0.48\textwidth}
        \centering
        \includegraphics[width=0.5\textwidth]{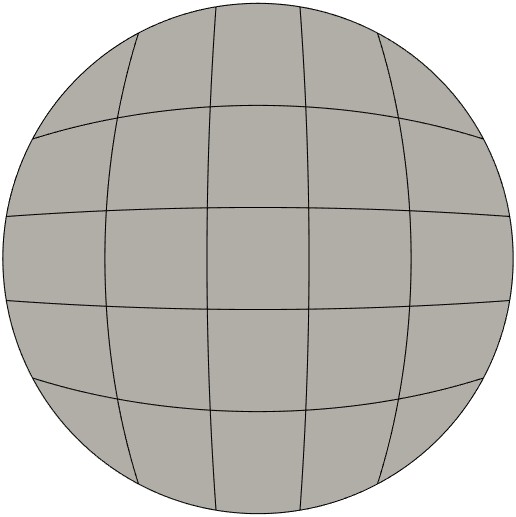}
        \caption{Isogeometric mesh of unit circle cross-section.}
        \label{fig:mesh_circle}
    \end{subfigure}

    \caption{Exemplary meshes for the two cross-sections: (a) square and (b) circle.}
    \label{fig:employed_meshes}
\end{figure*}

In this section, the present fully material, PK2 formulation of the CSWP is compared against results found in literature and against the analogous PK1 formulation from \cite{arora_computational_2019,herrnbock_PhdThesis_2023}. The solution $\mf u$ of the CSWP, as well as subsequent post-processed quantities, the beam stress resultants $\mf n, \mf m$ and stiffness $\mathbb{C}^b$, are compared and discussed.

For the results presented here, and without loss of generality, use of two exemplary cross-sections is made: a unit square of side length \SI{1.0}{mm} and a unit circle of radius \SI{1.0}{mm}. These are both parametrized using B-Splines as follows with 5 elements (knot spans) and degree $p=3$ in both parametric directions, see \cref{fig:employed_meshes}. 
Thus, there is a total of $n=64=(3+5)^2$ shape functions and control points for the discretization of the warping displacement, see \cref{Eq. Discr}.

Three different common hyperelastic material models are used for the constitutive relation, all available in the open source library \texttt{NLIGA} \cite{du_nliga_2020}: Saint Venant-Kirchhoff (SVK), Neo-Hooke (NH), and Mooney-Rivlin (MR). 
Relevant parameters, strain energy density formulations and modeling coefficients for the NH and MR approaches can be found within \cref{tab:materials}.

\begin{table}[tb]
    \centering
    \caption{Relevant strain energy density formulations and material parameters.}
    \begin{tabular}{c l c c}
        \toprule
        \multicolumn{4}{l}{Strain energy density formulations} \\
        \midrule
        \multicolumn{4}{l}{$\Bar{I}_1, \Bar{I}_2 \qquad ~\,: \text{First and second isochoric invariants of \textbf{C}}$} \\
        \addlinespace
        \multicolumn{4}{l}{$J = \text{det}(\mf F) : \text{Jacobian determinant of } \mf F$} \\
        \addlinespace

        \multicolumn{4}{l}{$\Psi_\text{SVK}(\mathbf{C}) = \frac{\lambda}{8} (\text{tr}(\mathbf{C} - \mathbf{I}))^2 + \frac{\mu}{4} \text{tr}((\mathbf{C} - \mathbf{I})^2)$} \\
        \addlinespace
        \multicolumn{4}{l}{$\Psi_\text{NH}(\mathbf{C}) = A_{10}(\Bar{I}_1 - 3) + \frac{K}{2}(J-1)^2$} \\
        \addlinespace
        \multicolumn{4}{l}{$\Psi_\text{MR}(\mathbf{C}) = B_{10}(\Bar{I}_1 - 3) + B_{01}(\Bar{I}_2-3)+\frac{K}{2}(J-1)^2$} \\
        \addlinespace
        \toprule
        Variable & Material parameter & Value & Unit \\
        \midrule
        $\lambda$ & 1st Lamé parameter & 121 & \SI{}{GPa} \\
        $\mu$ & 2nd Lamé parameter & 80 & \SI{}{GPa} \\
        $\nu$ & Poisson's ratio & 0.3 & -- \\
        $E$ & Young's modulus & 208.16 & \SI{}{GPa} \\
        $K$ & Bulk modulus & 174.34 & \SI{}{GPa} \\
        $A_{10}$ & NH coefficient ($\frac{\mu}{2}$) & 40 & $\SI{}{GPa}$ \\
        $B_{10}$ & MR coefficient 1 ($0.75 \cdot\frac{\mu}{2}$) & 30 & $\SI{}{GPa}$ \\
        $B_{01}$ & MR coefficient 2 ($0.25 \cdot\frac{\mu}{2}$) & 10 & $\SI{}{GPa}$ \\
        \bottomrule
    \end{tabular}
    \label{tab:materials}
\end{table}

For validation purposes, the PK1 formulation presented in \cite{herrnbock_PhdThesis_2023} was also implemented using \texttt{NLIGA} and was further used here as a means of comparison for the newly proposed formulation. Consequently, a thorough comparison between both formulations is made, yielding exactly the same results -- as shown in the following. 

In this section, for simplicity, the formulation from \cite{herrnbock_PhdThesis_2023,arora_computational_2019} is denoted as the ``PK1 formulation'', while the proposed fully material formulation in Voigt notation is denoted as the ``PK2 formulation''.

\paragraph{Comparison of residuals.}
A first indicator that both formulations are equivalent is the fact that they yield the same stiffness matrix $\hat{\mf K}$ and, consequently, the $L^2$-norm of the residual vector $\hat{\mf F}$ remains the same for both formulations in every Newton iteration, see \cref{tab:MultiAx_iterative_error} for an example with $\epsR = (0.02, 0.03, 0.1),\, \kR = (0.01,0.02,0.02)$,  applied to the unit square cross-section using the presented SVK material.

\begin{table}[tb]
    \centering
    \caption{$L^2$-norm of the residual vectors for PK1 and PK2 formulations in an exemplary loading with $\epsR = (0.02, 0.03, 0.1),\, \kR = (0.01,0.02,0.02)$.}
    \begin{tabular}{c S[table-format=1.2e-2] S[table-format=1.2e-2]}
        \toprule
        Newton step $k$ & {Residual (PK1)} & {Residual (PK2)} \\
        \midrule
        1 & {-} & {-} \\
        2 & 2.61e-1 & 2.61e-1 \\
        3 & 5.20e-6 & 5.20e-6 \\
        4 & 2.37e-14 & 2.37e-14 \\
        \bottomrule
    \end{tabular}
    \label{tab:MultiAx_iterative_error}
\end{table}

\begin{figure*}[t!]
    \centering

    \begin{subfigure}{0.48\textwidth}
        \centering
        \input{ValidationArora.tex}
        \caption{Comparison of torsional stiffness against literature \cite{arora_computational_2019} for a rectangular cross-section}
    \end{subfigure}
    \hfill
    \begin{subfigure}{0.48\textwidth}
        \centering
        \input{ValidationHerrnbock.tex}
        \caption{Comparison of torsional moment against literature \cite{herrnbock_PhdThesis_2023} for a square cross-section}
    \end{subfigure}

    \caption{Validation of torsional responses against literature references.}
    \label{fig:validation_Arora_HB}
\end{figure*}
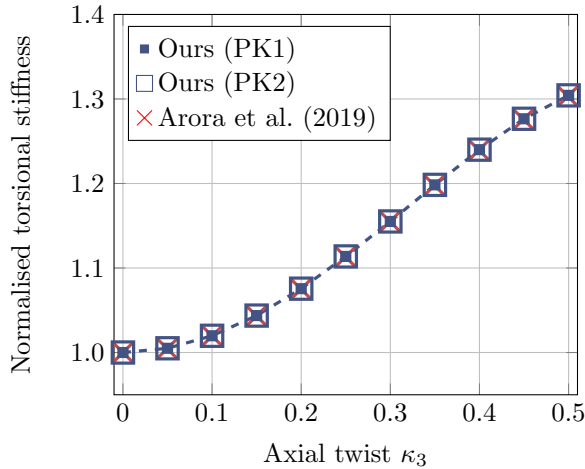
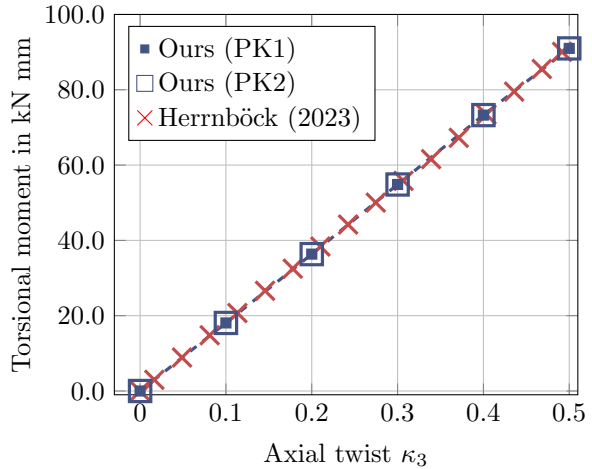

\paragraph{Pure torsion behavior.}
For validation of the implementation, a beam cross-section's reaction to torsion is compared to results found in the literature. 
Following \cite{arora_computational_2019}, a rectangular cross-section with aspect ratio $a/b=2$ is modeled using a SVK material with $\lambda=1.275$, $\mu=1.0$.
The dependency of the torsional stiffness on a prescribed axial twist $\kappa_3$ is compared to data from
\cite{arora_computational_2019}, see \cref{fig:validation_Arora_HB}. The normalized torsional stiffnesses in the present formulations exactly match the results from literature. 
Furthermore, a unit-square beam cross-section with SVK material and Lamé parameters $\lambda = 109\,995{,}8$ MPa and $\mu = 80\,194$ MPa from \cite{herrnbock_PhdThesis_2023} is investigated.
Here, the torsional moment $m_3$ over the axial twist $\kappa_3$ is evaluated against results presented in \cite{herrnbock_PhdThesis_2023}. Again, the results match exactly with the reference.

\begin{figure*}[h!]
    \centering
    \begin{subfigure}{0.48\textwidth}
        \centering
        \includegraphics[width=0.7\linewidth]{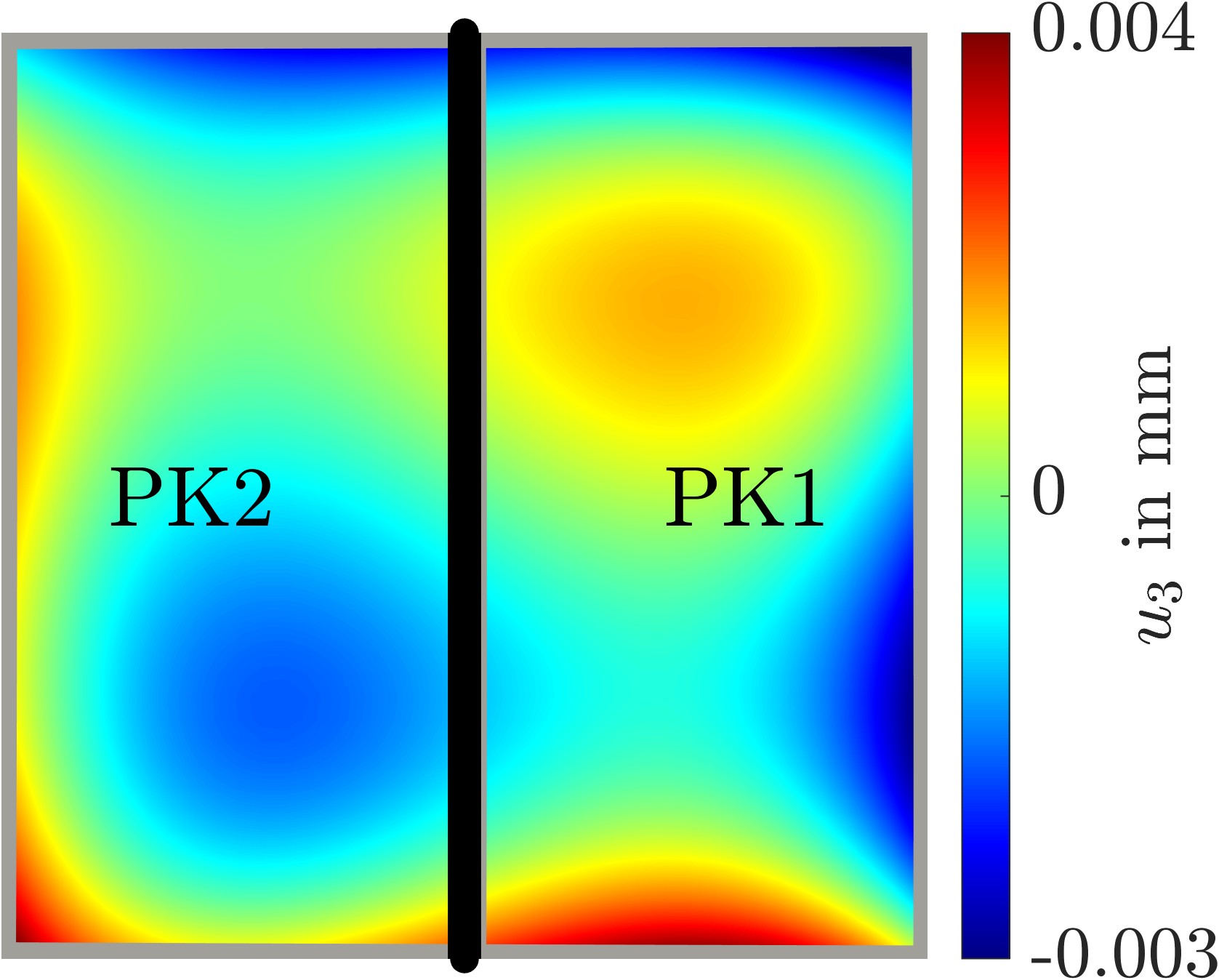}
        \caption{}
    \end{subfigure}
    \hfill
    \begin{subfigure}{0.48\textwidth}
        \centering
        \includegraphics[width=0.7\linewidth]{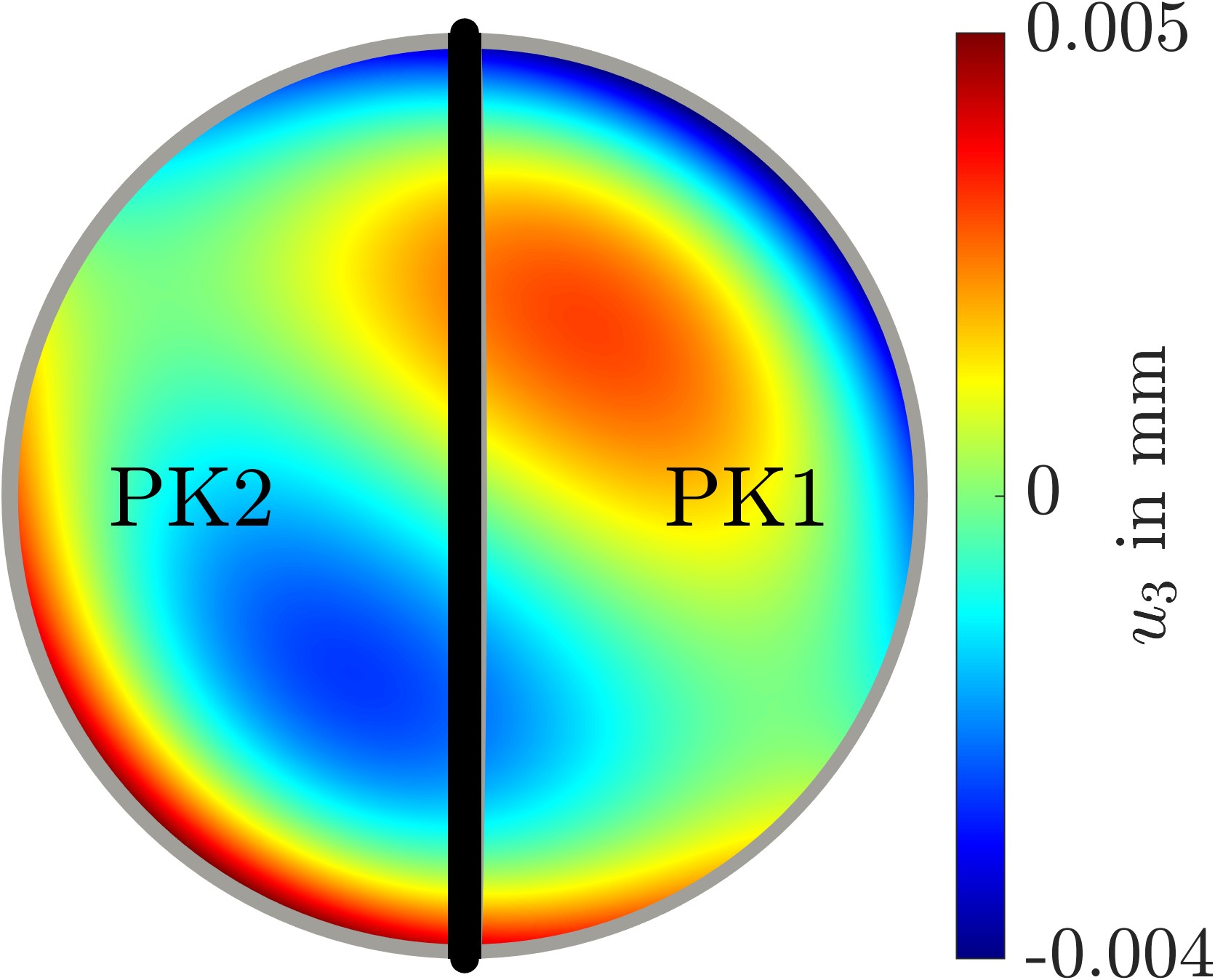}
        \caption{}
    \end{subfigure}
    \caption{Distribution of out-of-plane displacement $u_3$ for (a) a square  and (b) a circular cross-section under a multi-axial, stretch-dominated loading state. Solutions derived using the PK1 and PK2 formulations are separated by a black vertical line. The gray background indicates the undeformed cross-sections.}
    \label{fig:MultiAxial_U3}
\end{figure*}

\paragraph{Multi-axial loading behavior.}
For a multi-axial loading case dominated by axial strain with $\epsR = (0.02, 0.03, 0.1),\, \kR = (0.01,0.02,0.02)$, the warping displacement $\mf u$ is computed on the presented unit square and unit circle cross-sections using both PK1 and PK2 formulations with the SVK material. A color-gradient representation of the out-of-plane component $u_3$ is shown in \cref{fig:MultiAxial_U3}, highlighting the identical nature of the solution.

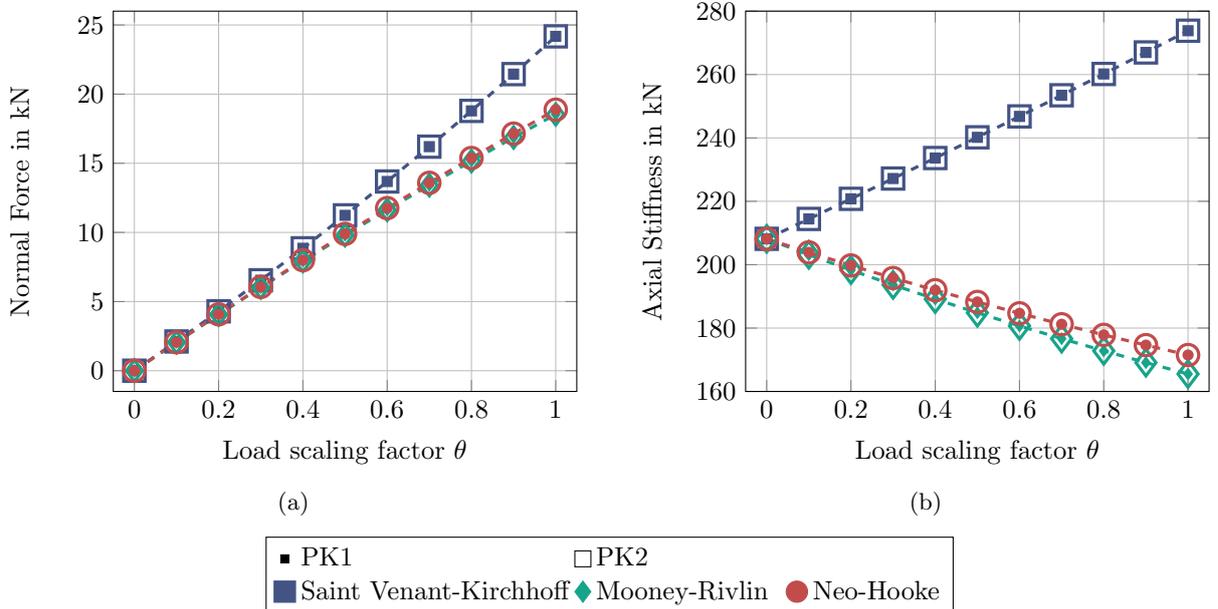
\begin{figure*}[tp!]
    \centering
    \begin{subfigure}{0.48\textwidth}
        \centering
        \input{PK_Comp_N3_Complex.tex}
        \caption{}
    \end{subfigure}
    \hfill
    \begin{subfigure}{0.48\textwidth}
        \centering
        \input{PK_Comp_C33_Complex.tex}
        \caption{}
    \end{subfigure}

    
    \ref*{common_legend}

    \caption{Comparison of responses for the different material models for (a) normal force and (b) axial stiffness as a function of the load scaling factor $\theta$ for the applied multi-axial, stretch-dominated loading on the unit-square cross-section.}
    \label{fig:MultiAx_N3_C33}
\end{figure*}

Post-processing quantities for the unit-square cross-section are also compared between both formulations. A linear scaling of the previously presented multi-axial loading case using the load scaling factor $\theta \in [0,1]$ is employed, where $\theta=0$ is no strain and $\theta = 1$ represents full loading. For each load step, the normal force and axial stiffness are compared between PK1 and PK2 formulations in \cref{fig:MultiAx_N3_C33} for all three material models described in \cref{tab:materials}. Again, it is visible that both PK1 and PK2 formulations match exactly, independent of the material model.

\begin{figure*}[tp!]
    \centering
    \begin{subfigure}{0.48\textwidth}
        \centering
        \includegraphics[width=0.7\linewidth]{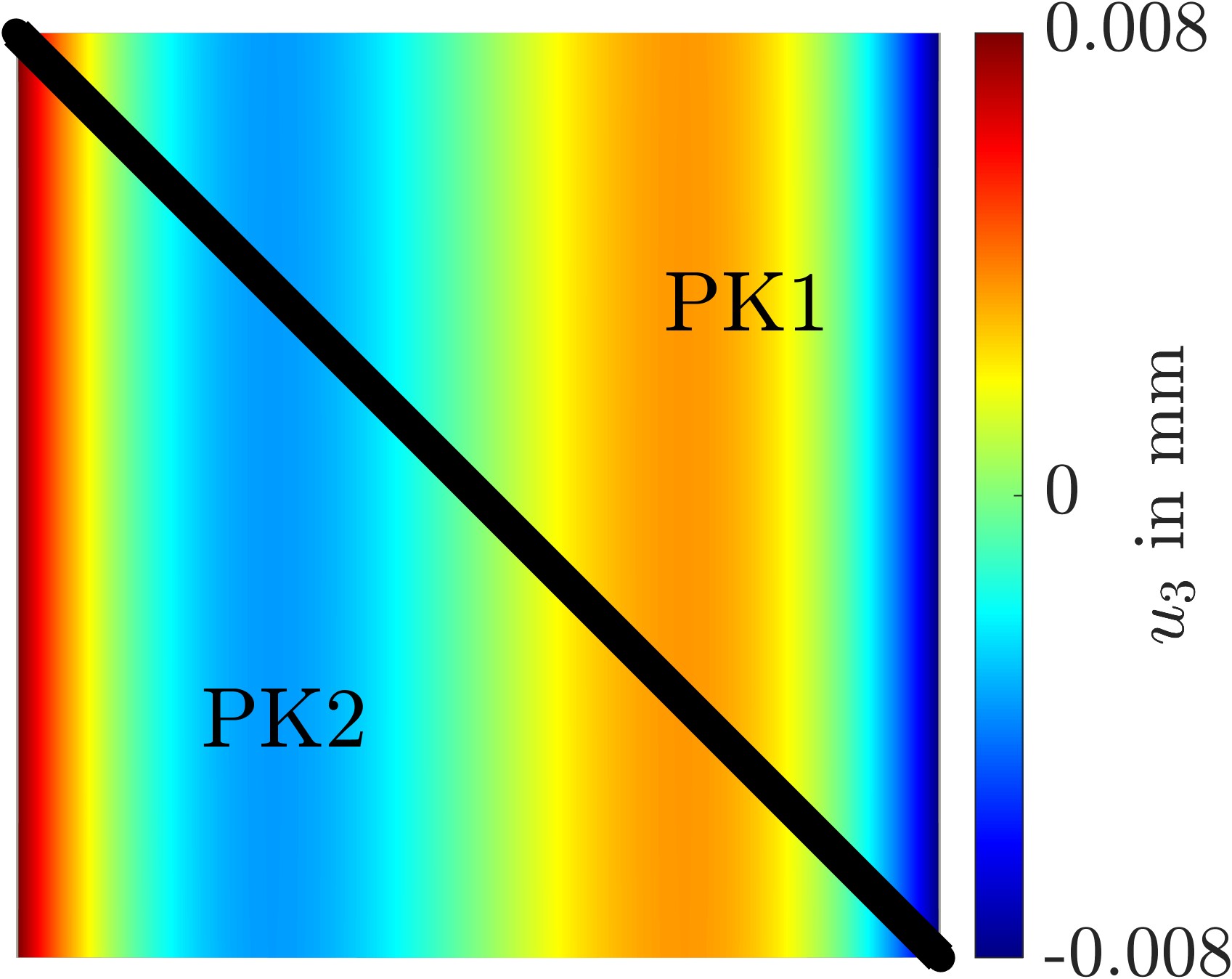}
        \caption{}
    \end{subfigure}
    \hfill
    \begin{subfigure}{0.48\textwidth}
        \centering
        \includegraphics[width=0.7\linewidth]{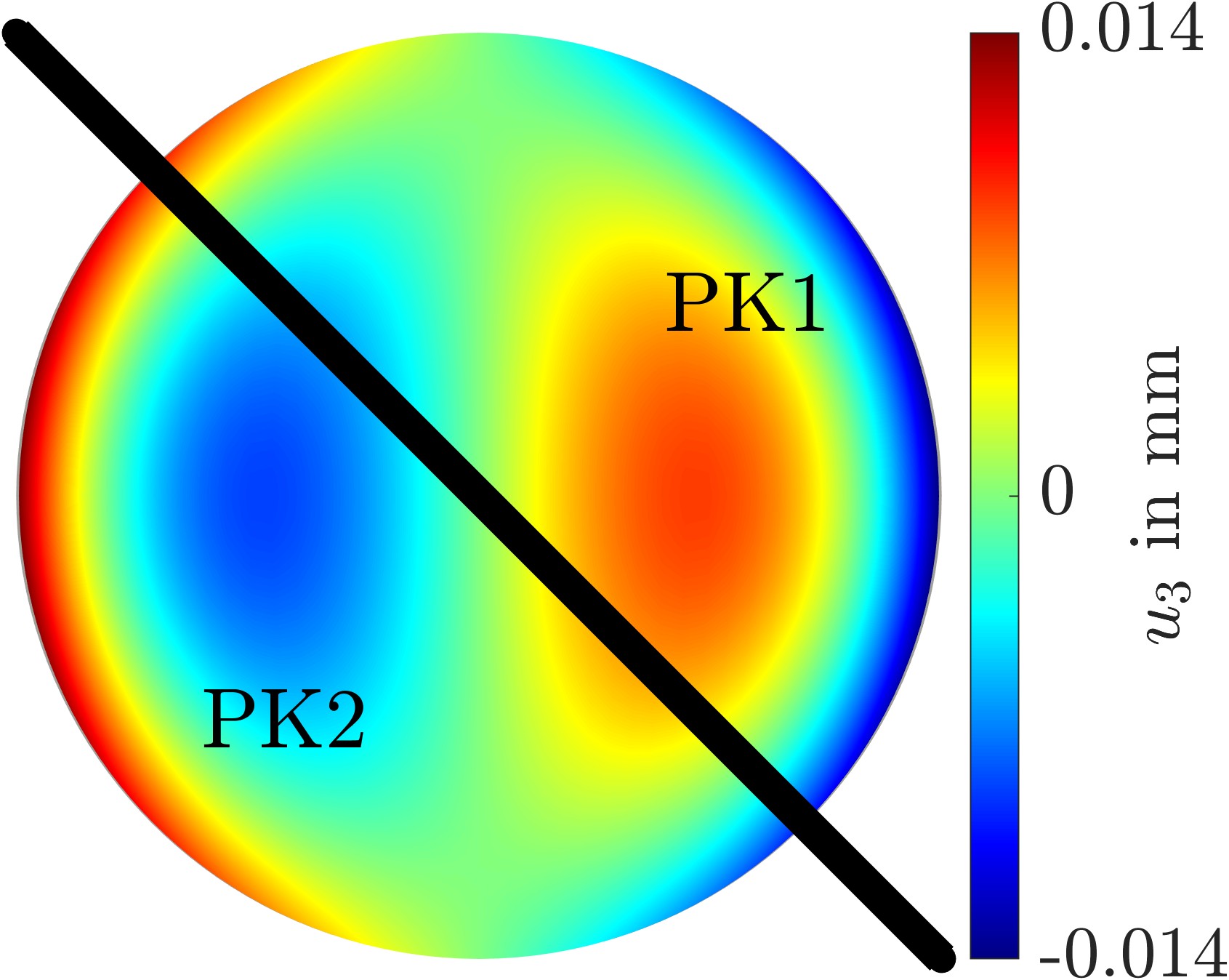}
        \caption{}
    \end{subfigure}

    \vspace{0.2cm}

    \begin{subfigure}{0.48\textwidth}
        \centering
        \includegraphics[width=0.7\linewidth]{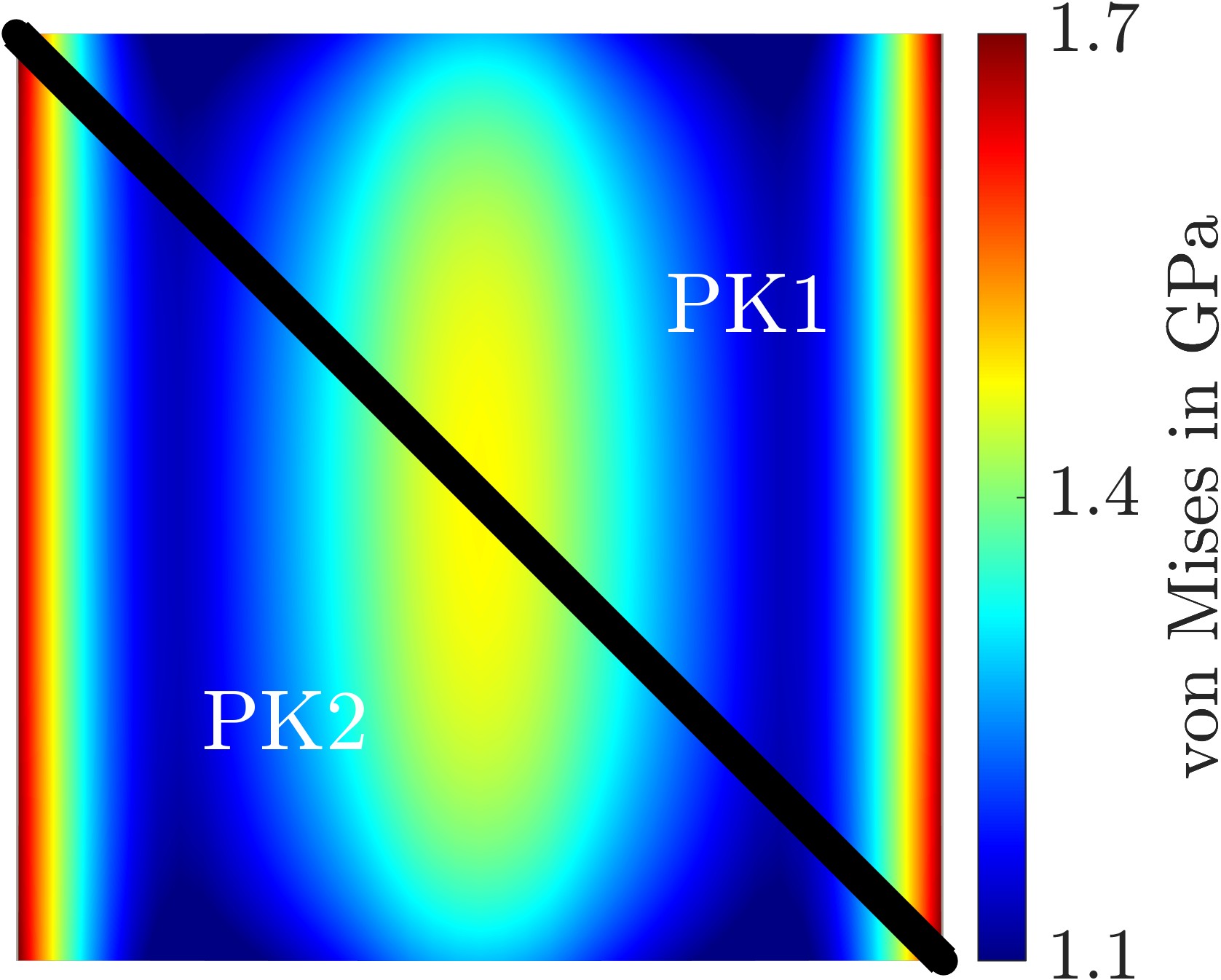}
        \caption{}
    \end{subfigure}
    \hfill
    \begin{subfigure}{0.48\textwidth}
        \centering
        \includegraphics[width=0.7\linewidth]{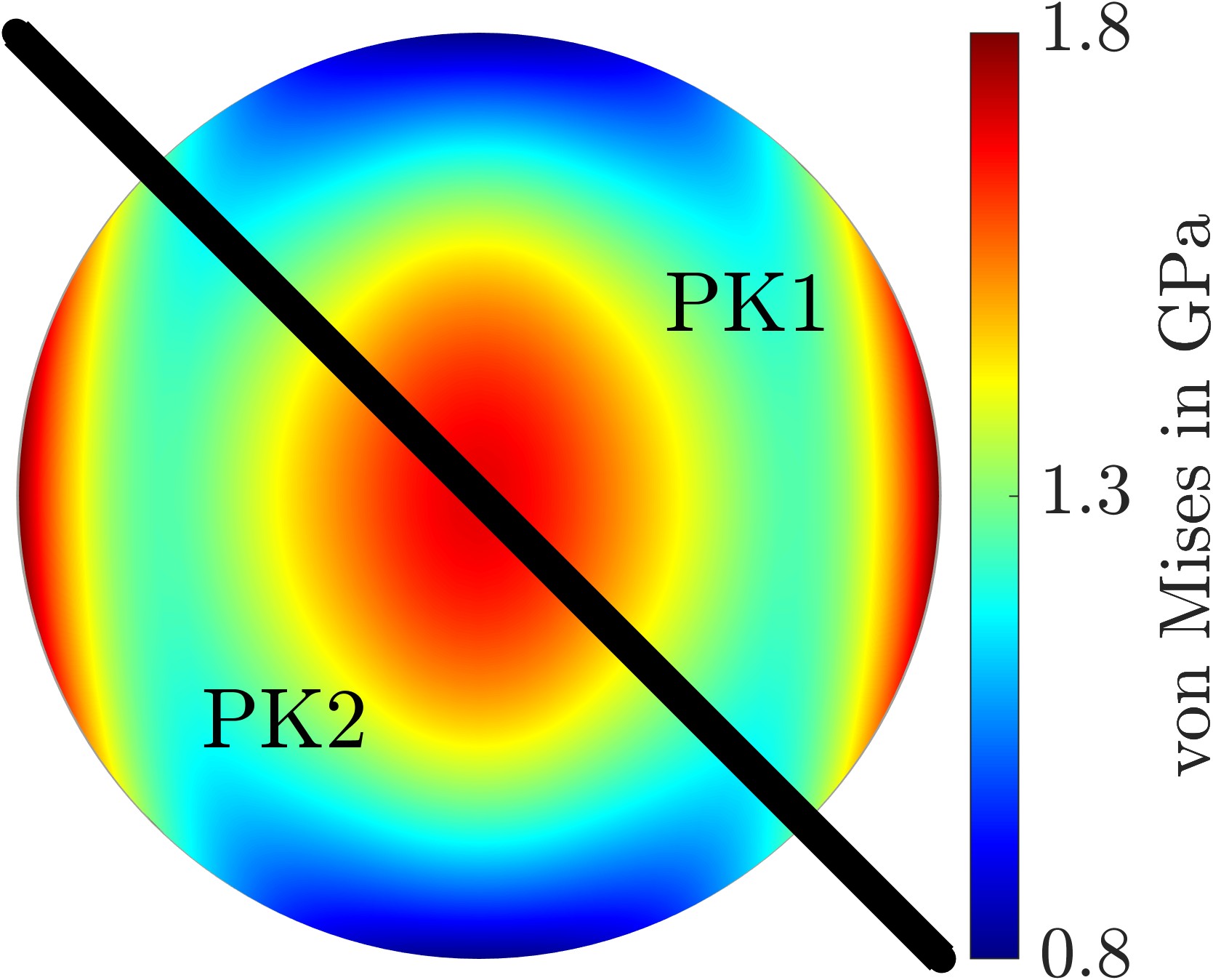}
        \caption{}
    \end{subfigure}
    \caption{Comparison between the distributions of out-of-plane Displacement $u_3$ (a, b) and the associated von Mises stresses (c, d) for a square and a circular cross-section under pure shear ($\epsR = [0.1, 0, 0]$).
    Solutions derived using the PK1 and PK2 formulations are separated by a black diagonal.}
    \label{fig:XBend_U3_VM}
\end{figure*}

\paragraph{Pure shear behavior.}
For a more visual analysis, a uniaxial pure shear load is investigated in \cref{fig:XBend_U3_VM} for the unit square and unit circle cross-sections with the SVK material presented in \cref{tab:materials}. Here, the out-of-plane displacement and the von Mises equivalent stress are visualized. 
We report that the out-of-plane deformation matches qualitatively the examples found in the literature \cite{arora_computational_2019,herrnbock_PhdThesis_2023}. Note that the circular cross-section exhibits a stress singularity in the von Mises stress, which is not displayed in \cref{fig:XBend_U3_VM}. This, however, is traced back to the singular isogeometric parametrization of a circle and not an issue related to the CSWP itself. Additional interactive visualizations of the results can be found in the corresponding \href{https://github.com/CPShub/CSWP_Voigt}{GitHub repository}.

%% file: ValidationArora.tex
\begin{tikzpicture}
    \begin{axis}[ 
        width=\linewidth,
        grid=both,
        xmin=-0.01,
        xmax=0.51, 
        xlabel={Axial twist $\kappa_3$},
        ymin=0.95,
        ymax=1.4, 
        ylabel={Normalised torsional stiffness},
        xtick={0,0.1,0.2,0.3,0.4,0.5},
        xticklabel style={
            /pgf/number format/.cd,
            fixed,
            precision=1,
            /tikz/.cd
        },
        yticklabel style={
            /pgf/number format/.cd,
            fixed,
            fixed zerofill,
            precision=1,
            /tikz/.cd
        },
        legend pos=north west,
        legend cell align={left},
    ]
        
        
        \addplot[forget plot, color=CPSred, solid, very thick,mark=x,only marks, mark size=5.0, fill=none] table[col sep=comma,x=k0_3,y=Arora_SVK] {PAPER_Validation_Arora.txt};
        
        \addplot[forget plot, color=CPSdarkblue, solid, very thick,mark=square*,only marks, mark size=1.5] table[col sep=comma,x=k0_3,y=c66_SVK_PK1] {PAPER_Validation_Arora_Us.txt};

        \addplot[forget plot, color=CPSdarkblue, dashed, very thick, no markers] table[col sep=comma,x=k0_3,y=c66_SVK_PK1] {PAPER_Validation_Arora_Us.txt};
        
        \addplot[forget plot, color=CPSdarkblue, solid, very thick,mark=square,only marks, mark size=4.0, fill=none] table[col sep=comma,x=k0_3,y=c66_SVK_PK2] {PAPER_Validation_Arora_Us.txt};


        \addlegendimage{only marks, mark=square*, mark size=1.5, color=CPSdarkblue}
        \addlegendentry{Ours (PK1)}
        
        \addlegendimage{only marks, mark=square, mark size=3.0, color=CPSdarkblue, fill=none}
        \addlegendentry{Ours (PK2)}

        \addlegendimage{only marks, mark=x, mark size=4, color=red}
        \addlegendentry{Arora et al. (2019)}

    \end{axis}
\end{tikzpicture}

%% file: ValidationHerrnbock.tex
\begin{tikzpicture}
    \begin{axis}[
        width=\linewidth,
        grid=both,
        xmin=-0.03,
        xmax=0.51, 
        xlabel={Axial twist $\kappa_3$},
        ymin=-1,
        ymax=100, 
        ylabel={Torsional moment in kN~mm},
        xtick={0,0.1,0.2,0.3,0.4,0.5},
        xticklabel style={
            /pgf/number format/.cd,
            fixed,
            precision=1,
            /tikz/.cd
        },
        yticklabel style={
            /pgf/number format/.cd,
            fixed,
            fixed zerofill,
            precision=1,
            /tikz/.cd
        },
        legend pos=north west,
        legend cell align={left},
    ]
        
        \addplot[forget plot, color=CPSred, solid, very thick,mark=x,only marks, mark size=5.0, fill=none] table[col sep=comma,x=k,y=M] {PAPER_Validation_Herrnbock.txt};
        
        \addplot[forget plot, color=CPSdarkblue, solid, very thick,mark=square*,only marks, mark size=1.5] table[col sep=comma,x=k0_3,y=m3_SVK_PK1] {PAPER_Validation_Herrnbock_Us.txt};

        \addplot[forget plot, color=CPSdarkblue, dashed, very thick, no markers] table[col sep=comma,x=k0_3,y=m3_SVK_PK1] {PAPER_Validation_Herrnbock_Us.txt};
        
        \addplot[forget plot, color=CPSdarkblue, solid, very thick,mark=square,only marks, mark size=4.0, fill=none] table[col sep=comma,x=k0_3,y=m3_SVK_PK2] {PAPER_Validation_Herrnbock_Us.txt};


        \addlegendimage{only marks, mark=square*, mark size=1.5, color=CPSdarkblue}
        \addlegendentry{Ours (PK1)}
        
        \addlegendimage{only marks, mark=square, mark size=3.0, color=CPSdarkblue, fill=none}
        \addlegendentry{Ours (PK2)}

        \addlegendimage{only marks, mark=x, mark size=4, color=red}
        \addlegendentry{Herrnböck (2023)}
 
    \end{axis}
\end{tikzpicture}

%% file: PK_Comp_N3_Complex.tex
\begin{tikzpicture}
    \begin{axis}[ 
						width=\linewidth,
                        grid=both,
						log basis x=2,
						xmin=-0.05,
						xmax=1.05, 
						xlabel={Load scaling factor $\theta$},
						ymin=-1.5,
						ymax=26, 
						ylabel={Normal Force in kN},
                        xtick={0,0.2,0.4,0.6,0.8,1.0},
                        xticklabel style={
                            /pgf/number format/.cd,
                            fixed,
                            precision=2,
                            /tikz/.cd
                        },
                        legend to name={common_legend},
                        legend columns=3, 
						legend style={
							cells={anchor=west},
							legend pos=north west, 
						}
					  ]
        \addplot[forget plot, color=CPSdarkblue, solid, very thick,mark=square*,only marks, mark size=1.5] table[col sep=comma,x=lambda,y=n3_SVK_pk1] {PAPER_Complex_Loading.txt};

        \addplot[forget plot, color=CPSdarkblue, dashed, very thick, no markers] table[col sep=comma,x=lambda,y=n3_SVK_pk1] {PAPER_Complex_Loading.txt}; 
        
        \addplot[forget plot, color=CPSdarkblue, solid, very thick,mark=square,only marks, mark size=4.0, fill=none] table[col sep=comma,x=lambda,y=n3_SVK_pk2] {PAPER_Complex_Loading.txt};

        \addplot[forget plot, color=CPSgreen, solid, very thick,mark=diamond*,only marks, mark size=1.5] table[col sep=comma,x=lambda,y=n3_MR_pk1] {PAPER_Complex_Loading.txt};

        \addplot[forget plot, color=CPSgreen, dashed, very thick, no markers] table[col sep=comma,x=lambda,y=n3_MR_pk1] {PAPER_Complex_Loading.txt};
        
        \addplot[forget plot, color=CPSgreen, solid, very thick,mark=diamond,only marks, mark size=4.0, fill=none] table[col sep=comma,x=lambda,y=n3_MR_pk2] {PAPER_Complex_Loading.txt};

        \addplot[forget plot, color=CPSred, solid, very thick,mark=*,only marks, mark size=1.5] table[col sep=comma,x=lambda,y=n3_NH_pk1]  {PAPER_Complex_Loading.txt};

        \addplot[forget plot, color=CPSred, dashed, very thick, no markers] table[col sep=comma,x=lambda,y=n3_NH_pk1]  {PAPER_Complex_Loading.txt};
        
        \addplot[forget plot, color=CPSred, solid, very thick,mark=o,only marks, mark size=4.0, fill=none] table[col sep=comma,x=lambda,y=n3_NH_pk2] {PAPER_Complex_Loading.txt};
    \end{axis}
\end{tikzpicture}

%% file: PK_Comp_C33_Complex.tex
\begin{tikzpicture}
    \begin{axis}[ 
						width=\linewidth,
                        grid=both,
						log basis x=2,
						xmin=-0.05,
						xmax=1.05, 
						xlabel={Load scaling factor $\theta$},
						ymin=160,
						ymax=280, 
						ylabel={Axial Stiffness in kN},
                        xtick={0,0.2,0.4,0.6,0.8,1.0},
                        xticklabel style={
                            /pgf/number format/.cd,
                            fixed,
                            precision=2,
                            /tikz/.cd
                        },
                        legend to name={common_legend},
                        legend columns=3, 
						legend style={
							cells={anchor=west},
							legend pos=north west, 
						}
					  ]
        \addplot[forget plot, color=CPSdarkblue, solid, very thick,mark=square*,only marks, mark size=1.5] table[col sep=comma,x=lambda,y=C33_SVK_pk1] {PAPER_Complex_Loading.txt};

        \addplot[forget plot, color=CPSdarkblue, dashed, very thick, no markers] table[col sep=comma,x=lambda,y=C33_SVK_pk1] {PAPER_Complex_Loading.txt};
        
        \addplot[forget plot, color=CPSdarkblue, solid, very thick,mark=square,only marks, mark size=4.0, fill=none] table[col sep=comma,x=lambda,y=C33_SVK_pk2] {PAPER_Complex_Loading.txt};

        \addplot[forget plot, color=CPSgreen, solid, very thick,mark=diamond*,only marks, mark size=1.5] table[col sep=comma,x=lambda,y=C33_MR_pk1] {PAPER_Complex_Loading.txt};

        \addplot[forget plot, color=CPSgreen, dashed, very thick, no markers] table[col sep=comma,x=lambda,y=C33_MR_pk1] {PAPER_Complex_Loading.txt};
        
        \addplot[forget plot, color=CPSgreen, solid, very thick,mark=diamond,only marks, mark size=5, fill=none] table[col sep=comma,x=lambda,y=C33_MR_pk2] {PAPER_Complex_Loading.txt};

        \addplot[forget plot, color=CPSred, solid, very thick,mark=*,only marks, mark size=1.5] table[col sep=comma,x=lambda,y=C33_NH_pk1] {PAPER_Complex_Loading.txt};

        \addplot[forget plot, color=CPSred, dashed, very thick, no markers] table[col sep=comma,x=lambda,y=C33_NH_pk1] {PAPER_Complex_Loading.txt};
        
        \addplot[forget plot, color=CPSred, solid, very thick,mark=o,only marks, mark size=4.0, fill=none] table[col sep=comma,x=lambda,y=C33_NH_pk2] {PAPER_Complex_Loading.txt};

        \addlegendimage{only marks, mark=square*, mark size=1.5, color=black}
        \addlegendentry{PK1}
        
        \addlegendimage{only marks, mark=square, mark size=3.0, color=black, fill=none}
        \addlegendentry{PK2}

        \addlegendimage{empty legend}
        \addlegendentry{}

        \addlegendimage{only marks, mark=square*, mark size=4, color=CPSdarkblue}
        \addlegendentry{Saint Venant-Kirchhoff}

        \addlegendimage{only marks, mark=diamond*, mark size=4, color=CPSgreen}
        \addlegendentry{Mooney-Rivlin}

        \addlegendimage{only marks, mark=*, mark size=4, color=CPSred}
        \addlegendentry{Neo-Hooke}
    \end{axis}
\end{tikzpicture}

%% file: S6Conclusion.tex
\section{Conclusion}
\label{Sec. Concl}

In this contribution, we reinterpret the cross-sectional warping problem for hyperelastic beams and propose a fully material formulation based on the PK2 stress and GL strain tensors. The proposed formulation exploits their symmetric structure, enabling the use of Voigt notation and thereby reducing fourth-order elasticity tensors and second-order stress tensors to a matrix--vector representation. This leads, by design, to improved computational efficiency. To formulate the CSWP in Voigt notation, the classical $\mf B$- and $\mathfrak{B}$-operators of nonlinear FEM are modified to exactly represent the cross-sectional deformation and its associated deformation gradient in terms of the strain measures of the beam theory.

The proposed formulation is verified by reproducing results available in the literature. In addition, we implement the corresponding PK1 formulation of \cite{arora_computational_2019,herrnbock_PhdThesis_2023} and compare both approaches under multi-axial loading. The resulting forces, moments, stiffnesses, and iteration errors are found to be identical in both formulations.

To promote reproducibility, the implementations of both formulations, together with all numerical results, are made available in a \href{https://github.com/CPShub/CSWP_Voigt}{GitHub repository}. This may serve to promote broader adoption and further extension of the proposed formulation to related problems involving nonlinear beam constitutive models.

Owing to its close connection to classical nonlinear finite element formulations, the present approach can be naturally extended to more complex settings, such as plasticity \cite{eidel_elastoplastic_2003} and multiphysics problems, e.g., chemoelasticity \cite{shafqat_ChemMechSBeam_2025}. 
For geometrically complex cross-sections, further gains in computational efficiency may be achieved, for instance, by formulating the CSWP as an immersed problem, as demonstrated by Elbadry et al. \cite{elbadryImmersedIGA2024} within a general 2D hyperelasticity framework using \texttt{NLIGA}.